\documentclass[letterpaper,twocolumn,10pt]{article}
\usepackage{usenix2019_v3}

 \usepackage{nopageno}

\usepackage{amsmath}

\usepackage{tikz}
\usepackage{subcaption}
\usetikzlibrary{shapes.geometric, arrows, positioning}
\usepackage{amsfonts, amsmath,amssymb}
\usetikzlibrary{arrows}
\usepackage{float}

\usepackage{tikz}
\usepackage{xspace}
\usepackage[utf8]{inputenc}
\usepackage{amsmath}
\usepackage{cite}
\usepackage{tabularx}
\usepackage{multirow}
\usepackage{makecell}
\usepackage{booktabs}
\usepackage{graphicx,wrapfig,lipsum}
\usepackage[linesnumbered, noend, ruled,vlined]{algorithm2e}
\usepackage{chngcntr}
\usepackage{hyperref, url}
\usepackage{epstopdf}
\usepackage{soul}
\newcommand{\ie}{{i.e.,~}}
\newcommand{\eg}{{e.g.,~}}
\newcommand{\etal}{{et~al.}} 
\newcommand{\sota}{the state-of-the-art }
\newcommand{\T}[1]{\textit{\text{'#1'}}}
\newcommand{\TT}[1]{``\textit{#1}"}
\newcommand\myeq{\mkern1.5mu{=}\mkern1.5mu}

\newcommand{\nsure}[1]{I'LL WRITE IT BETTER: \textcolor{orange}{#1}}

\newcommand{\RS}{R}
\newcommand{\Xgt}{{X}_{A}}
\newcommand{\Xd}{W}
\newcommand{\af}{\pi}
\newcommand{\AS}{\mathbf{X}}
\newcommand{\apply}[1]{(#1)}
\newcommand{\WS}{\mathbb{W}}
\newcommand{\dw}{dictionary-word}
\newcommand{\dws}{dictionary-words}
\newcommand{\adam}{\textit{AdaMs}}

\begin{document}
	%
	\title{Reducing Bias in Modeling Real-world Password Strength\\ via Deep Learning and Dynamic Dictionaries }

	\author{
		{\rm Dario Pasquini}$^{\dag,\S}$, 
		{\rm Marco Cianfriglia}$^{\S}$,  
		{\rm Giuseppe Ateniese}$^{\ddag}$ and 
		{\rm Massimo Bernaschi}$^{\S}$\\
		$^{\dag}$\textit{Sapienza University of Rome}, 
		$^{\ddag}$\textit{Stevens Institute of Technology}, 
		$^{\S}$\textit{Institute of Applied Computing CNR}
	}

	\maketitle
	
	\renewcommand{\thefootnote}{\fnsymbol{footnote}}
	\footnotetext[1]{This paper appears in the proceedings of the 30th USENIX Security Symposium 2021.}%
	
	\renewcommand{\thefootnote}{\arabic{footnote}}
	
	\begin{abstract}
Password security hinges on an in-depth understanding of the techniques adopted by attackers. Unfortunately, real-world adversaries resort to pragmatic guessing strategies such as dictionary attacks that are inherently difficult to model in password security studies. In order to be representative of the actual threat, dictionary attacks must be thoughtfully configured and tuned. However, this process requires a domain-knowledge and expertise that cannot be easily replicated. 
The consequence of inaccurately calibrating dictionary attacks is the unreliability of password security analyses, impaired by a severe~measurement~bias.

		In the present work, we introduce a new generation of dictionary attacks that is consistently more resilient to inadequate configurations. Requiring no supervision or domain-knowledge, this technique automatically approximates the advanced guessing strategies adopted by real-world attackers. To achieve this: (1) We use deep neural networks to model the proficiency of adversaries in building attack configurations. (2)~Then, we introduce dynamic guessing strategies within dictionary attacks. These mimic experts' ability to adapt their guessing strategies on the fly by incorporating knowledge on their targets.\par

		Our techniques enable more robust and sound password strength estimates within dictionary attacks, eventually reducing overestimation in modeling real-world threats in password security.
	\end{abstract}
	\section{Introduction}
	Passwords have proven to be irreplaceable. They are still preferred over safer options and appear essential in fallback mechanisms. However, users tend to select their passwords as easy-to-remember strings, which results in very skewed distributions that an attacker can easily model. This makes passwords and authentication systems that implement them inherently susceptible to guessing attacks. 
In this scenario, the security of the authentication protocol cannot be stated via a security parameter (e.g., the key size).
The only way to establish the soundness of a system is to model adversarial behaviors and cast accurate adversary models. To this end, simulating password guessing attacks has become a pivotal task.
\par

In this direction, more than three decades of active research provided us with powerful password models \cite{MM, PCFG, FLA, imprl}. However, very little progress has been made to systematically model real-world attackers and their guessing strategies~\cite{measuring, reasoning}. As a matter of fact, \textit{password crackers} rarely harness fully-automated approaches developed in academia. They rely on more pragmatic guessing techniques that present stronger inductive biases. In offline attacks, experts use high-throughput, and extremely flexible techniques such as \textbf{dictionary attacks with mangling rules}~\cite{unix}. This class of attacks produces candidate passwords by expanding a dictionary/wordlist through a set of scripted string transformations (a rules-set) which aim at mimicking users' composition habits such as \textit{leeting} (\eg \TT{pa\$\$w0rd}) or concatenating digits (\eg \TT{password123})~\cite{the_benefit}.
\par

	Unlike fully-automated approaches, dictionary attacks are heavily sensitive to their initial configuration. To be effective, these must rely on highly tuned setups---pairs of dictionaries and mangling rules-sets that have been carefully optimized and thoroughly calibrated. To cast such configurations, real-world attackers rely on a manual process that is based on specific expertise that can only be achieved and refined over years of practical experience~\cite{ars}. Furthermore, attackers customize their configurations for the current target by dynamically adjusting the dictionary and rules-set leveraging information gathered before or during the attack.
\par

Unfortunately, lacking the same domain-knowledge of experts, most researchers and security practitioners performing dictionary attacks in their security analysis rely on \textit{off-the-shelf} setups and static guessing strategies that only remotely approximate the actual effectiveness of real-world attacks. Indeed, as demonstrated in~\cite{measuring}, these commonly used default configurations bring to a profound overestimation of password strength that fails to correctly approximate adversarial capabilities. Unavoidably, this introduces a strong \textbf{bias} in the produced strength estimates that fundamentally sways the conclusion of security analysis.\\ 

In the present paper, we move towards reducing this inherent measurement bias by devising a new generation of dictionary attacks that automates the advanced guessing strategies adopted by attackers; we cast an adversary model that is consistently more resilient to inaccurate configurations, and that better describes real-world attackers' capabilities. To that purpose, we introduce general procedures that systematically mimic different adversarial behaviors:

First, by relying on deep learning techniques, we devise the \textbf{Adaptive Mangling Rules attack}. This artificially simulates the optimal configurations harnessed by expert adversaries by explicitly handling \textbf{the conditional nature of mangling rules}. Here, during the attack, each word from the dictionary is associated with a dedicated and possibly unique rules-set created at runtime via a \textbf{deep neural network.}
Using this technique, we confirmed that standard attacks, based on \textit{off-the-shelf} dictionaries and rules-sets, are sub-optimal and can be easily compressed up to an order of magnitude in the number of guesses. 
Furthermore, we are the first to explicitly model the strong relationship that binds mangling rules and dictionary words, demonstrating its connection with optimal configurations in dictionary attacks.

Then, we introduce {\bf dynamic guessing strategies} within dictionary attacks~\cite{imprl}. Real-world adversaries perform their guessing attacks incorporating prior knowledge on the targets and dynamically adjusting their guesses during the attack. In doing so, professionals seek to optimize their configurations and maximize the number of compromised passwords. Unfortunately, automatic guessing techniques fail to model this adversarial behavior. Instead, we demonstrate that dynamic guessing strategies can be enabled in dictionary attacks and substantially improve the guessing attack's effectiveness even without prior optimization. More prominently, our technique makes dictionary attacks consistently more resilient to misconfigurations by promoting the completeness of the dictionary at runtime.\\

Finally, we combine these general methodologies and introduce the Adaptive Dynamic Mangling rules attack (\adam). The \adam~attack consistently reduces the overestimation induced by sub-optimal configurations in dictionary attacks, enabling more reliable and sound password strength estimates.
	\paragraph{Organization:} Section~\ref{sec:back} gives an overview of the fundamental concepts needed for the comprehension of our contributions. In Section~\ref{sec:aff_and_amr}, we introduce Adaptive Mangling Rules aside the intuitions and tools on which those are based. Section~\ref{sec:dynamic} discusses dynamic mangling rules attacks. Finally, Section~\ref{sec:adam}~aggregates the previous methodologies, introducing the \adam~attack. The  motivation and evaluation of the proposed techniques are presented in their respective sections. Section~\ref{conclusion} concludes the paper, although supplementary information is provided in the Appendices.
	\section{Background and preliminaries}
	\label{sec:back}
	We start by covering password guessing attacks and their foundations in Section~\ref{sec:pg}. In Section \ref{sec:dictattack}, we focus on dictionary attacks that are the basis of our contributions. Next, Section \ref{sec:related} briefly discusses relevant related works. Finally, we define the threat model in Section~\ref{sec:tc}.
\subsection{Password Guessing}
\label{sec:pg}
Human-chosen passwords do not distribute uniformly in the exponentially large key-space. Users tend to choose easy-to-remember passwords that aggregate in relatively few dense clusters.
Real-world passwords, therefore, tend to cluster in very bounded distributions that can be modeled by an attacker, making authentication-systems intrinsically susceptible to \textbf{guessing attacks}. In a guessing attack, the attacker aims at recovering plaintext credentials by attempting several candidate passwords (guesses) till success or \textit{budget} exhaustion; this happens by either searching for collisions of password hashes (\textbf{offline attack}) or attempting remote logins (\textbf{online attack}). In this process, the attacker relies on a so-called \textbf{password model} that defines which, and in which order, guesses should be tried to maximize the effectiveness of the attack (see Section~\ref{sec:tc}).
\par
Generally speaking, a password model can be understood as a suitable estimation of the password distribution that enables an educated exploration of the key-space. Existing password models construct over a heterogeneous set of assumptions and rely on either intuitive or rigorous security definitions. From the most practical point of view, those can be divided into two macro-classes: parametric and nonparametric password models.
\par
Parametric approaches build on top of probabilistic reasoning; they assume that real-world password distributions are sufficiently smooth to be accurately described from suitable parametric probabilistic models. Here, a password mass function is explicitly \cite{FLA, MM} or implicitly \cite{imprl} derived from a set of observable data (\ie previously leaked passwords) and used to assign a probability to each element of the key-space. During the guessing attack, guesses are produced by traversing the key-space following the decreasing probability order imposed by the modeled mass function. These approaches are, in general, relatively slow and unsuitable for practical offline attacks. Although simple models such as Markov Chains can be employed \cite{jtr_mark}, more advanced and effective models such as the neural network ones \cite{FLA, imprl} are hardly considered outside the research domain due to their inefficiency.
\par

Nonparametric models such as Probabilistic Context-Free Grammars (PCFG) and dictionary attacks rely on simpler and more intuitive constructions, which tend to be closer to human logic. Generally, those assume passwords as realizations of templates and generate novel guesses by abstracting and applying such patterns on ground-truth. These approaches maintain a collection of tokens that are either directly given as part of the model configuration (\eg the dictionary and rules-set for dictionary attack.) or extracted from observed passwords in a setup phase (\eg terminals/grammar for PCFG). In contrast with parametric models, these can produce only a limited number of guesses, which is a function of the chosen configuration. A detailed discussion on dictionary attacks follows in the next section.

\subsection{Dictionary Attacks}
\begin{table}[t]
	\centering
	\resizebox{.9\columnwidth}{!}{%
		\begin{tabular}{c|c|p{5cm}}
			\large \textbf{Rule} & \large \textbf{Result} & \large \textbf{Rule description}.\\
			\toprule
			\texttt{r} & \TT{niemtel} & Reverse string.\\
			\midrule
			\texttt{T0} & \TT{Letmein} & Capitalize the first character.\\
			\midrule
			\texttt{\$9 \$9} & \TT{letmein99} & Append \TT{99} to the string.\\
			\midrule
			\texttt{se3} & \TT{l3tm3in} & Substitute the character \T{e} with \T{3}.\\
			\midrule
			\texttt{] ] ] \$m \$a \$n} & \TT{letmman} & Remove the last three symbols and append the string \TT{man}.\\
			\bottomrule
		\end{tabular}
	}
	\caption{Example of mangling rules and their effect on the dictionary-word~\TT{letmein}. The rules are selected from the rules-set \textit{Best64}.}
	\label{tab:mag_examples}
\end{table}
\label{sec:dictattack}
Dictionary attacks can be traced back to the inception of password security studies \cite{worm,unix}. They stem from the observation that users tend to pick their passwords from a bounded and predictable pool of candidates; common natural words and numeric patterns dominate most of this skewed distribution~\cite{pop}. An attacker, collecting such strings (\ie creating a dictionary/wordlist), can use them as high-quality guesses during a guessing attack, rapidly covering the key-space's densest zone. These dictionaries are typically constructed by aggregating passwords revealed in previous incidents and plain-word dictionaries.
\par

Although dictionary attacks can produce only a limited number of guesses\footnote{The required disk space inherently bounds the number of guesses issued from plain dictionary attacks. Guessing attacks can quickly go beyond $10^{12}$ guesses, and storing such a quantity of strings is not practical.}, these can be extended through \textbf{mangling rules}. Mangling rules attacks describe password distributions by factorizing guesses in two main components: (1)~dictionary-words and (2)~string transformations (mangling rules). These transformations aim at replicating users' composition behaviors. Mangling transformations are modeled by the attacker and collected in sets (rules-sets). During the guessing attack, each dictionary word is extended in real-time through mangling rules, creating novel guesses that augment the guessing attack's coverage over the key-space. Hereafter, we use the terms dictionary attack and mangling rules attack interchangeably.
\par

 Most widely known implementations of mangling rules are included in the \textit{password cracking} software \textit{Hashcat}~\cite{HC} and \textit{John the Ripper}~\cite{jtr} (JtR). Here, mangling rules are encoded through simple custom programming languages. Table~\ref{tab:mag_examples} reports some instances of mangling rules and their effect. \textit{Hashcat} and \textit{JtR} share almost overlapping mangling rules languages, although few peculiar instructions are unique to each tool. However, they consistently differ in the way mangling rules are applied during the attack. \textit{Hashcat} follows a \textbf{word-major order}, where all the rule-set rules are applied to a single dictionary-word before the next dictionary word is considered. In contrast, \textit{JtR} follows a \textbf{rule-major order}, where a rule is applied to all the dictionary words before moving to the next rule. In our work, we rely on the approach of \textit{Hashcat} as the word-major order is necessary to efficiently implement the adaptive mangling rules attack that we introduce in Section~\ref{sec:amra}.\par
 
 The community behind these software packages developed numerous mangling rules sets that have been made public. Such sets have a heterogeneous size and can range between tens to thousands of entries. Mangling rules can be either manually crafted by human experts and optimized through public competitions \cite{crackme} or produced via simple automatic procedures \cite{auto_mr}. Here, it is important to note that public rules-sets are often sub-optimal when compared to highly-tuned, private sets harnessed by experts~\cite{reasoning}.
 \par
 
 Despite their simplicity, mangling rules attacks represent a substantial threat in offline password guessing. Mangling rules are swift and inherently parallel; they are naturally suited for both parallel hardware (\ie GPUs) and distributed setups, making them one of the few guessing approaches suitable for large-scale attacks (\eg botnets).\par
 
Furthermore, real-world attackers update their guessing strategy dynamically during the attack~\cite{measuring}. Basing on prior knowledge and the initially matched passwords, they tune their guesses generation process to describe their target set of passwords better and eventually recover more of them. To this end, professionals prefer extremely flexible tools that allow for fast and complete customization. While \sota probabilistic models fail at that, dictionary attacks make any form of customization feasible as well as natural.
 
\subsection{Related Works}
\label{sec:related}
Although dictionary attacks are ubiquitous in password security research \cite{meter, eco, persuasion, FLA, composition_policy}, little effort has been spent studying them. This section covers the most relevant contributions.
\par

Ur~\etal~\cite{measuring} firstly made explicit the large performance gap between optimized and stock configurations for mangling rules attacks. In their work, Ur~\etal~recruited professional figures in password recovery and compared their performance against \textit{off-the-shelf} parametric/nonparametric approaches in different guessing scenarios. Here, professional attackers have been shown capable of vastly outperform any password model. This thanks to custom dictionaries, proprietary mangling rules, and the ability to create tailored rules for the attacked set of passwords.
Finally, the authors show that the performance gap between professional and non-professional attackers can be reduced by combining guesses of multiple password models.
\par

More recently, Liu~\etal~\cite{reasoning} produced a set of tools that can be used to optimize the configuration of dictionaries attacks. These solutions extend previous approaches \cite{auto_mr, rule_hc}, making them faster. Their core contribution is an algorithm capable of inverting almost all mangling rules; that is, given a rule $r$ and password to evaluate $p$, the inversion-rule function produces as output a \textit{regex} that matches all the preimages of $r(p)$ \ie all the dictionary entries that transformed by $r$ would produce $p$. At the cost of an initial pre-computation phase, following this
approach, it is possible to count dictionary-words/mangling-rules \textit{hits} (\ie guessed passwords) on an attacked set without enumerating all the possible guesses. Liu~\etal~used the method to optimize the ordering of mangling rules in a rules-set by sorting them in decreasing hits-count order.\footnote{Primarily, for rule-major order setups (\eg JtR).} In doing so, the authors observed that default rules-sets follow an optimal ordering only rarely.
\par 

Basing on the same general approach, they speedup the automatic generation of mangling rules \cite{auto_mr} and augment dictionaries by adding missing words in consideration of known attacked sets \cite{rule_hc}. Similarly, they derive an approximate guess-number calculator for rule-major order attacks.
\subsection{Threat Model}
\begin{table}[t]
	\centering
	\resizebox{.9\columnwidth}{!}{%
		\begin{tabular}{c|c|p{5cm}}
			\textbf{Name} & \textbf{\makecell{Unique\\Passwords}} & \textbf{Brief Description}\\ \toprule
			\textit{LinkedIn}\cite{link} & $60.599.259$ & \small An employment-oriented online service. \\\midrule
			\textit{youku}~\cite{youku} & $47.487.499$ & \small Chinese video hosting service.\\\midrule
			\textit{MyHeritage}\cite{myhe} & $36.393.972$ & \small Online genealogy platform.\\\midrule
			\textit{zooks}\cite{zooks}& $29.010.979$ & \small Online dating service available.\\\midrule
			\textit{RockYou}\cite{rock} & $14.344.391$ & \small Gaming platform. \\\midrule
			\textit{animoto}\cite{animoto} & $8.420.466$ & \small A cloud-based video creation service.\\\midrule
			\textit{zomato}\cite{zomato} & $4.955.821$ & \small Indian, food delivery application. About $40\%$ of the password are random tokens of six alphanumeric characters. \\ \midrule
			\textit{phpBB} & $184.389$ & \small Software website.\\
			\bottomrule
		\end{tabular}
	}
	\caption{Password leaks used in the paper sorted by size.}
	\label{tab:password}
\end{table}
\label{sec:tc}
In our study, we primarily model the case of trawling, offline attacks. Here, an adversary aims at recovering a set of passwords $\mathbf{X}$ (also referred to as \textit{attacked-set}) coming from an arbitrary password distribution $P(\mathbf{x})$ by performing a guessing attack. To better describe both the current trend in password storing techniques \cite{bcrypt,pkcs, mem_hard} and real-world attackers' goals~\cite{economics_pg}, we assume a rational attacker who is bound to produce a limited number of guesses. 
More precisely, this attacker aims at maximizing the number of guessed passwords in $\mathbf{X}$ given a predefined \textit{budget} \ie a maximal number of guesses the attacker is willing to perform on $\mathbf{X}$. Hereafter, we model this strategy under the form of $\beta$-\textit{success-rate} \cite{entropy, the_science_of_guessing}:
\begin{equation}
\label{eq:bsr}
s_{\beta}(X) = \sum_{i=1}^{\beta}P(x_i).
\end{equation}
\paragraph{Experimental setup}
In our construction, we do not impose any limitation on the nature of $P(\mathbf{x})$ nor the attacker's \textit{a priori} knowledge. However, in our experiments, we consider a weak attacker who does not retain any initial knowledge of the target distribution \ie who cannot provide an optimal attack configuration for $\mathbf{X}$ before the attack. This last assumption
makes a better description of the use-case of automatic guessing approaches currently used in password security studies.
\par

In the attacks reported in the paper, we always sort the words in the dictionary according to their frequency. Additionally, in the reported results for all the dictionary attacks, we do not count guesses that remain unchanged after the application of a mangling rule ($r(w) = w$). This aims to avoid biases in measuring the effectiveness of the adaptive approach presented in Section~\ref{sec:amra}. 
The password leaks that we use through the paper are listed in Table~\ref{tab:password}.
	\section{The Adaptive Mangling Rules attack}
	\label{sec:aff_and_amr}
\begin{figure*}[h!]
	\captionsetup[subfigure]{justification=centering}
	\centering
	\resizebox{1\textwidth}{!}{%
		
		\begin{subfigure}{.26\textwidth}\includegraphics[scale=.4]{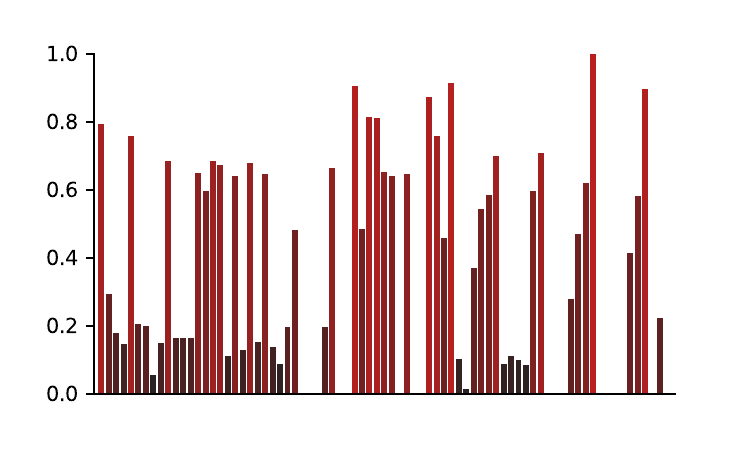}\caption{Only digits.}\label{fig:hits0}\end{subfigure}
		\begin{subfigure}{.26\textwidth}\includegraphics[scale=.4]{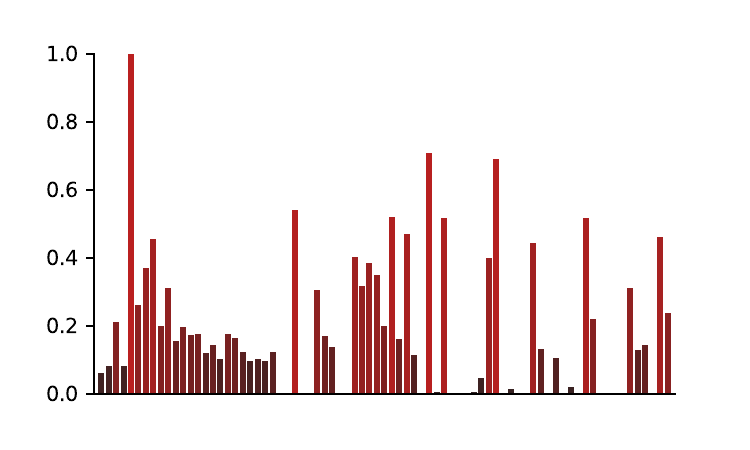}\caption{Only capital letters.}\label{fig:hits1}\end{subfigure}
		\begin{subfigure}{.26\textwidth}\includegraphics[scale=.4]{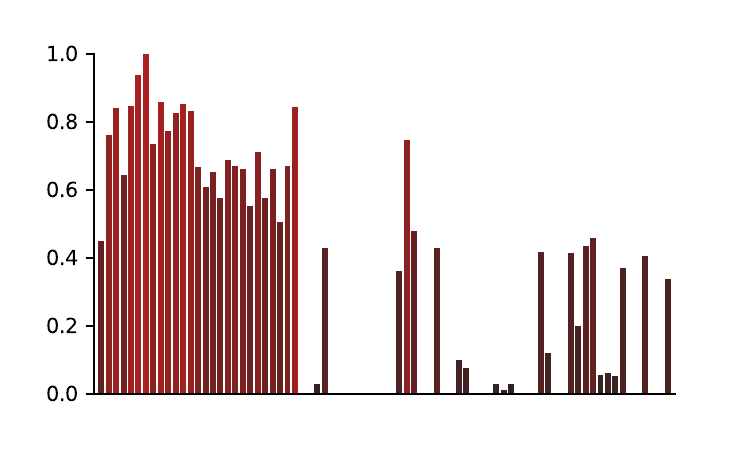}\caption{Strings of length $5$.}\label{fig:hits2}\end{subfigure}
		\begin{subfigure}{.26\textwidth}\includegraphics[scale=.4]{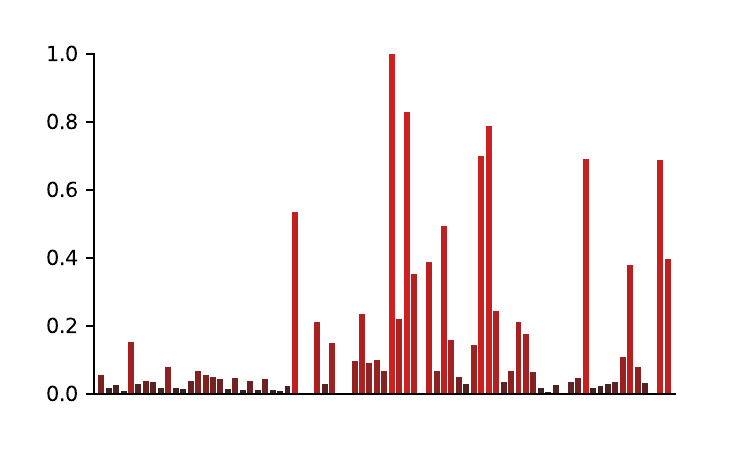}\caption{Strings of length $10$.}\label{fig:hits3}\end{subfigure}
		
	}
	\caption{Distribution of hits per rule for $4$ different input dictionaries for the same attacked-set \ie \textit{animoto}. Within a plot, each bar depicts the normalized number of hits for one of the $77$ mangling rules in \textit{best64}. We performed the attack with \textit{Hashcat}.}
	\label{figure:meter_out_ex}
\end{figure*}
This section introduces the first core block of our password model: the Adaptive Mangling Rules. 
We start in Section~\ref{sec:fcd}, where we make explicit the conditional nature of mangling rules while discussing its connection with optimal attack configurations. In Section~\ref{sec:affinity}, we model the functional relationship connecting mangling rules and dictionary words via a deep neural network. Finally, leveraging the introduced tools, we establish the Adaptive Mangling Rules attack in Section~\ref{sec:amra}.
\paragraph{Motivation:} 
Dictionary attacks are highly sensitive to their configuration; while parametric approaches tend to be more robust to training sets and hyper-parameters choices, the performance of dictionary attacks crucially depends on the selected dictionary and rules-set~\cite{measuring, reasoning}. As evidenced by Ur~\etal~\cite{measuring}, real-world attackers rely on extremely optimized configurations. Here, dictionaries and mangling rules are jointly created over time through practical experience \cite{ars}, harnessing a domain knowledge and expertise that is mostly unknown to the academic community~\cite{reasoning}.
\par

Password security studies often rely on publicly available dictionaries and rules-sets that are not as effective as advanced configurations adopted by professionals. Unavoidably, this leads to a constant overestimation of password strength that skews studies and reactive analysis conclusions.
\par

Hereafter, we show that professional attackers' domain-knowledge can be suitably approximated with a \textbf{Deep Neural Network}. Given that, we devise a new dictionary attack that autonomously promotes functional interaction between the dictionary and the rules-set, implicitly simulating the precision of real-world attackers' configurations.\\ 
We start by presenting the intuition behind our technique. Formalization and methodology are reported later.
\subsection{The conditional nature of mangling rules}
\label{sec:fcd}
As introduced in Section~\ref{sec:dictattack}, dictionary attacks describe password distributions by factorizing guesses into two main components---a dictionary word~$w$ and a transformation rule~$r$. Here, the word~$w$ acts as a \textbf{semantic base}, whereas~$r$ is a \textbf{syntactic transformation} that aims at providing a suitable guess through the manipulation of~$w$. Generally speaking, such factorized representation can be thought of as an approximation of the typical users' composition behavior: starting from a plain word or phrase, users manipulate it by performing operations such as \textit{leeting}, appending characters or concatenation.\par

At configuration time, such transformations are abstracted and collected in arbitrary large rules-sets under the form of mangling rules. Then, during the attack, guesses are reproduced by exhaustively applying the collected rules to all the dictionary words. \textbf{In this generation process, rules are applied unconditionally on all the words, assuming that the abstracted syntactic transformations equally interact with all the dictionary elements.}\par

However, arguably, users do not follow the same simplistic model in their password composition process. Users first select words and then mangling transformations conditioned by those words. That is, mangling transformations are subjective and depend on the base words on which those are applied. For instance, users may prefer to append digits at the end of a name (\eg \TT{jimmy} to \TT{jimmy91}), repeat short words rather than long ones (\eg \TT{why} to \TT{whywhywhy}) or capitalize certain strings over others (\eg \TT{cookie} to \TT{COOKIE}). A similar intuition was harnessed in~\cite{sempcfg}, where the semantic of words was considered in defining context-free grammars for passwords.
%

In this direction, we can think of each mangling rule as a function that is valid on an arbitrary small subset of the dictionary space, strictly defined by the users' composition habits. 
Thus, applying a mangling rule on words outside this domain unavoidably brings it to produce guesses that have only a negligible probability of inducing hits during the guessing attack (\ie that do not replicate users' behavior).
This concept is captured in Figure~\ref{figure:meter_out_ex}, where four panels depict the \textit{hits} distribution of the rules-set \TT{\textit{best64}} for four different dictionaries. Each dictionary represents a specific subset of the dictionary space that has been obtained by filtering out suitable strings from the \textit{RockYou} leak; namely, these are passwords composed of: digits (Figure~\ref{fig:hits0}), capital letters (Figure~\ref{fig:hits1}), passwords of length $5$ (Figure~\ref{fig:hits2}), and passwords of length $10$ (Figure~\ref{fig:hits3}).
The four histograms show how mangling rules selectively and heterogeneously interact with the underlying dictionaries. Rules that produce many hits for a specific dictionary inevitably perform very poorly with the others. 
\par

Eventually, the conditional nature of mangling rules has a critical impact in defining the effectiveness of a dictionary attack. To reach optimal performance, an attacker has to resort to a setup that \textit{a priori} maximizes the conditional effectiveness of mangling rules.
In this direction, we can see highly optimized configurations used by experts as pairs of dictionaries and rules-sets that organically support each other in the guesses generation process.\footnote{This has also been indirectly observed by Ur~\etal~in their ablation study on pro's guessing strategy, where the most remarkable improvement was achieved with a proprietary dictionary in tandem with a proprietary rules-set.}
On the other hand, configurations based on arbitrary chosen rule-sets and dictionaries may not be fully compatible, and, as we show later in the paper, they generate many low-quality guesses. Unavoidably, this phenomenon makes adversary models based on mangling rules inaccurate and induce an overestimation of password strength \cite{measuring}.\\

Next, we show how modeling the conditional nature of mangling rules allows us to cast dictionary attacks that are inherently more resilient to poor configurations. 

\subsection{A Model of Rule/Word Compatibility}
\label{sec:affinity}
We introduce the notion of \textbf{compatibility} that refers to the functional relation among dictionary words and mangling rules discussed in the previous section.
The compatibility can be thought of as a continuous value defined between a mangling rule $r$ and a \dw~$w$ that, intuitively, measures the \textit{utility} of applying the rule $r$ on $w$. More formally, we model {\em compatibility} as a function: \[\af:\RS \times \WS \rightarrow [0,1],\] where $\RS$ and $\WS$ are the rule-space (\ie the set of all the suitable transformations $r:\WS \rightarrow \WS$) and the dictionary-space (\ie the set of all possible dictionary words), respectively. Values of $\af(w, r)$ close to $1$ indicate that the transformation induced by $r$ is well-defined on $w$ and would lead to a valuable guess. Values close to $0$, instead, indicate that users would not apply $r$ over $w$, \ie guesses will likely fall outside the dense zone of the password distribution.
\par
This formalization of the compatibility function also leads to a straightforward probabilistic interpretation that better supports the learning process through a neural network. Indeed, we can think of $\af$ as a probability function over the event:
 \[r(w)\in \AS,\]
where $\AS$ is an attacked set of passwords. More precisely, we have that:
\[\forall_{w\in \WS,\ r \in \RS} \big(\af(r, w) = P(r(w)\in \AS) \big).
\]
In other words, $P(r(w)\in \AS)$ is the probability of guessing an element of $\AS$ by trying the guess $g=r(w)$ produced by the application of $r$ over $w$. Furthermore, such a probability can be seen as an unnormalized version of the password distribution, creating a direct link to probabilistic password models~\cite{FLA, MM} as we have that:
\[\small
\forall_{w\in \WS,\ r \in \RS} \langle \frac{\af(r, w)}{Z} = P(r(w)) \rangle \]
for an intractable partition function $Z$. This follows from the observation that:
\[
	\small
	\begin{array}{c}
		\forall g_i,g_j \in \mathbb{X} : P(g_i) \ge P(g_j) \Leftrightarrow P(r_i\apply{x_i} \in \AS)\ge P(r_j\apply{x_j} \in \AS) \\ 
		\text{with}: \ g_i=r_i\apply{x_i}\ \text{and}\ g_j=r_j\apply{x_j},
	\end{array}
\]
where $\mathbb{X}$ is the key-space. However, this password distribution is defined over the factorized domain $\RS\times \WS$ rather than directly over the key-space. This factorized form offers us practical advantages over the classic formulation. More in detail, by choosing and fixing a specific rule-space $\RS$ (\ie a rules-set), we can reshape the compatibility function as:
\begin{equation}
 \af_\RS: \WS \rightarrow [0,1]^{|\RS|}.
 \label{eq:reshaped_aff}
\end{equation}
This version of the compatibility function takes as input a \dw~and outputs a compatibility value for each rule in the chosen rule-set with\textbf{ a single inference}. This form is concretely more computational convenient and will be used to model the neural approximation of the compatibility function.

Next, we show how the compatibility function can be inferred from raw data using deep learning.

\subsubsection{Learning the compatibility function}
\label{sec:learning_aff}
As stated before, the probabilistic interpretation of the compatibility function makes it possible to learn $\af$ using a neural network. Indeed, the probability $P(r\apply{w}\in \AS)$, in any form, can be described through a binary classification. That is, {for each pair word/rule ($w$, $r$), we have to predict one of two possible outcomes: $g\in\AS$ or $g\not\in\AS$, where $g=r(w)$.} In solving this classification task, we can train a neural network in a logistic regression and obtain a good approximation of the probability~$P(r\apply{w}\in \AS)$.
\par
In the same way, the reshaped formulation of $\af$ (\ie Eq.~\ref{eq:reshaped_aff}) describes a \textbf{multi-label classification}. In multi-label classification, each input participates simultaneously to multiple binary classifications; an input is associated with multiple classes at the same time.
More formally, having a fixed number of possible classes $n$, each data point is mapped to a binary vector in $\{0,1\}^{n}$. In our case, $n=|\RS|$ and each bit in the binary vector corresponds to the outcome of the event $r_j(w)\in X$ for a rule $r_j\in R$.
\par

To train a model, then, we have to resort to a supervised learning approach. We have to create a suitable training-set composed of pairs \textit{(input,label)} that the neural network can model during the training. Under our construction, we can easily produce such suitable labels by performing a mangling rules attack. In particular, fixed a rules-set $\RS$, we collect pairs $(w_i, y_i)$, where $w_i$ is the input to our model (\ie a \dw) and $y_i$ is the label vector associated with $w_i$. As explicated before, the label $y_i$ asserts the membership of the list of guesses $[r_1(w_i), r_2(w_i), \dots, r_{|R|}(w_i)]$ over a hypothetical target set of passwords $\AS$: 
\begin{equation}
	\label{eq:label}
	y_i=[r_1(w_i)\in \AS,\ r_2(w_i)\in \AS,\ \dots,\ r_{|R|}(w_i)\in \AS]
\end{equation}
To collect labels, we have to concertize $\AS$ by choosing a representative set of passwords. 
Intuitively, such a set should be as large and diverse as possible as it aims at describing the entire key-space. Hereafter, we refer to this set as $\Xgt$. This is the set of passwords we attack during the process of collecting labels. Similarly, we have to choose another set of strings $\Xd$ that represents the dictionary-space. This is used as input to the neural network during the training process and as the input dictionary during the simulated guessing attack. Details on the adopted set are given at the end of the section.
\par
Finally, given $\Xgt$ and $\Xd$, and chosen a rules-space $\RS$, we construct the set of labels by simulating a guessing attack; that is, for each entry $w_i$ in the dictionary $\Xd$, we collect the label vector $y_i$ (E.q.~\ref{eq:label}). In doing so, we used a modified version of \textit{Hashcat} described in Appendix~\ref{app:imple}. Alternatively, the technique proposed in~\cite{reasoning} can be used to speed up the collection of the labels.
\par 
Unlike the actual guessing attack, in the process, we do not remove passwords from $\Xgt$ when those are guessed correctly; that is, the same password can be guessed multiple times by different combinations of rules and words. This is necessary to correctly model the functional compatibility. In the same way, we do not consider the identity mangling rule (\ie $\T{:}$) in the construction of the training set. When it occurs, we remove it from the rules set. To the same end, we do not consider hits caused by \textit{conditional identity transformations} \ie $r\apply{w}=w$. 
\paragraph{Training set configuration}
The creation of a training set entails the proper selection of the sets $\Xgt$ and $\Xd$ as well as the rules-set $\RS$. Arguably, the most critical choice is the set $\Xgt$, as this is the ground-truth on which we base the approximation of the compatibility function. In our study, we select $\Xgt$ to be the password leak discovered by \textit{4iQ} in the \textit{Dark Web} \cite{4iq}. 
We completely anonymized all entries by removing users' information and obtained a set of $\sim 4 \cdot10^8$ of {\em unique} passwords. We use this set as $\Xgt$ within our models.\\
Similarly, we want $\Xd$ to be a good description of the dictionary-space. However, in this case, we are supported by the generalization capability of the neural network that can automatically obtain a more general description of the input space. In our experiments, we use the \textit{LinkedIn} leak as~$\Xd$.
\par

Finally, we train three neural networks that learn the compatibility function for three different rules-sets; namely \textit{PasswordPro}, \textit{generated} and \textit{generated2}. Those sets are provided with the \textit{Hashcat} software and widely studied in previous works \cite{reasoning, imprl, FLA}. Table \ref{tab:rules} lists them along with some additional information.
\par

Eventually, the labels we collect in the guessing process are extremely sparse. In our experiments, more than $95\%$ of the guesses are a miss, causing our training-set to be extremely unbalanced towards the negative class.
\begin{table}[t]
	\centering
	\resizebox{.8\columnwidth}{!}{%
		\begin{tabular}{c|c|p{5cm}}
			\textbf{Name} & \textbf{Cardinality} & \textbf{Brief Description}\\ \toprule
			\textit{PasswordPro} & $3120$ & Manually produced.\\ \midrule
			\textit{generated} & $14728$ & Automatically generated.\\ \midrule
			\textit{generated2} & $65117$ & Automatically generated.\\ 
			\bottomrule
		\end{tabular}
	}
 \caption{Used \textit{Hashcat}'s mangling rules sets.}
	\label{tab:rules}
\end{table}
\paragraph{Model definition and training}
\label{sec:model_and_train}
We construct our model over a residual structure \cite{resnet} primarily composed of mono-dimensional convolution layers. Here, input strings are first embedded at character-level via an embedding matrix; then, a series of residual blocks are sequentially applied to extract a global representation for dictionary words. Finally, such representations are mapped into the label-space by means of a single, linear layer that performs the classification task. To note that, although the model applies over sequential data, the use of a convolutional network instead of a recurrent one is essential to reduce inference latency. This will be critical in the context of our application (see Section~\ref{sec:amra}).
\par

This architecture is trained in a multi-label classification; each output of the final dense layer is squashed in the interval $[0,1]$ via the sigmoid function, and binary cross entropy is applied to each probability separately. The network's loss is then obtained by summing up all the cross-entropies of the $|\RS|$ classes/rules.
\par
As mentioned in the previous section, our training-set is extremely unbalanced toward the negative class; that is, the vast majority of the ground-truth labels assigned to a training instance are negative (\ie the application of a rule on the word does not bring to a hit). Additionally, a similar disproportion appears in the distribution per rule. Typically, we have many rules that count only a few positive examples, whereas others have orders of magnitude more hits. In our framework, we alleviate the negative effects of those disproportions by inductive bias. In particular, we achieve it by considering a \textbf{focal regulation} in our loss function~\cite{focalloss}.
\par
Originally developed for object detection tasks in which there is a strong imbalance between foreground and background classes, we adopt focal regulation to account for sparse and underrepresented labels when learning the compatibility function. This \textit{focal loss} is mainly characterized by a modulating factor~$\gamma$ that dynamically reduces the importance of well-classified instances in the computation of the loss function, allowing the model to focus on hard examples (\eg underrepresented rules). More formally, the form of regularized binary cross entropy that we adopt is defined as:
\[
	\text{FL}(p_j, y_j) = \begin{cases}
	-(1-\alpha)(1-p_j)^\gamma \log(p_j) & \text{if } y_j=1 \\
	\alpha p_j^\gamma \log(1 - p_j) & \text{if } y_j=0 \end{cases},
\]
where $p_j$ is the probability assigned by the model to the $j$'th class, and $y_j$ is the ground-truth label (\ie 1/hit and 0/miss). The parameter~$\alpha$ in the equation allows us to declare an \textit{a~priori} importance factor to the negative class. We use that to down-weighting the correct predictions of the negative class in the loss function that would be dominant otherwise. In our setup, we dynamically select~$\alpha$ based on the distribution of the hits observed in the training set. In particular, we choose~$
\alpha\myeq\frac{\bar{p}}{(1-\bar{p})}$, where~$\bar{p}$ is the ratio of positive labels (\ie hits/guesses) in the dataset. Differently, we fix~$\gamma\myeq2$ as we found this value to perform well via empirical evaluation.
\par

Summing up, our loss function is defined as:
\[
\mathcal{L}_f = \mathbb{E}_{x, y} \sum_{j=1}^{|\RS|} \text{FL}( \textit{sigmoid}(f(x)_j), y_j)
\]
where $f$ are the \textit{logits} of the neural network. We train the model using \textit{Adam} stochastic gradient descent~\cite{adam} until an early-stopping-criteria based on the \textit{AUC} computed on a validation set is reached.
\par

Maintaining the same general architecture, we train different networks with different sizes. In our experiments, we noticed that large networks provide a better approximation of the compatibility function, although small networks can be used to reduce the computational cost with a limited loss in utility. This suggests that modeling compatibility between rules and words is complex and that simpler models with less capacity (e.g., not based on deep neural networks) should perform poorly. In the paper, we report the results only for our biggest networks.\par

We implemented our framework on \textit{TensorFlow}; the models have been trained on a \textit{NVIDIA DGX-2} machine. A complete description of the architectures employed is given in Appendix~\ref{app:arch}. 
\par

Ultimately, we obtain three different neural networks: one for each rule-set reported in Table~\ref{tab:rules}. 
The suitability of these neural approximations will be proven later in the paper.
\paragraph{Additional approaches} To improve the performance of our method, we further investigated domain-specific constructions for multi-label classification. In particular, we tested label embedding techniques together we deep architectures. Those are approaches that aim at modeling, implicitly, the correlation among labels. However, although unconditional dependence is evident in the modeled domain, we found no concrete advantage in considering it during the training. In the same direction, we investigated more sophisticated embedding techniques, where labels and dictionary-words were jointly mapped to the same latent space~\cite{C2AE}, yet achieving similar or worse performance.
\par
Additionally, we tested implementations based on \textit{transformer} networks \cite{attention}, obtaining no substantial improvement. We attribute such a result to the lack of dominant long-term relationships among characters composing dictionary-words. In such a domain, we believe convolutional filters to be fully capable of capturing characters' interactions. Furthermore, convolutional layers are significantly more efficient than the multi-head attention mechanism used by \textit{transformer} networks.
%
	\subsection{Adaptive Mangling Rules}
	\label{sec:amra}
\begin{figure*}[t!]
	\centering
	\includegraphics[trim = 0mm 87mm 0mm 0mm, clip, width=.4\linewidth]{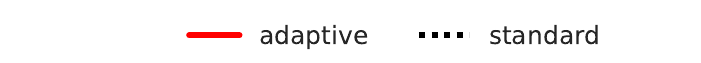}\\
	\begin{subfigure}{.22\textwidth}
		\centering
		\includegraphics[trim = 0mm 0mm 0mm 0mm, clip, width=1\linewidth]{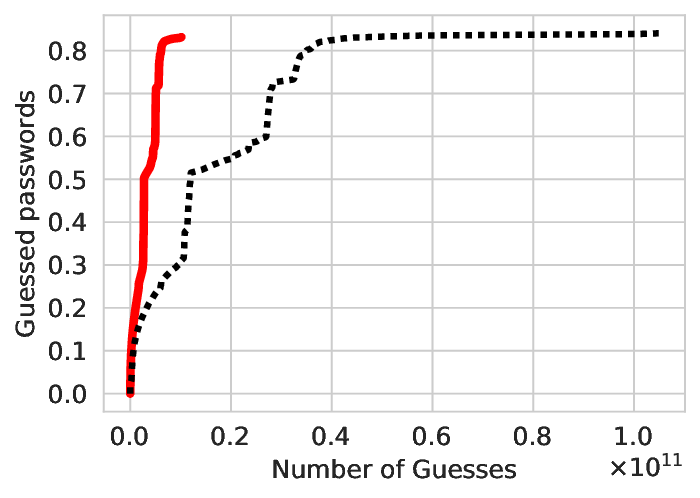}
		\caption{\textit{MyHeritage} on \textit{animoto}}\label{fig:ad_vs_cl_a}
	\end{subfigure}\begin{subfigure}{.22\textwidth}
		\centering
		\includegraphics[trim = 0mm 0mm 0mm 0mm, clip, width=1\linewidth]{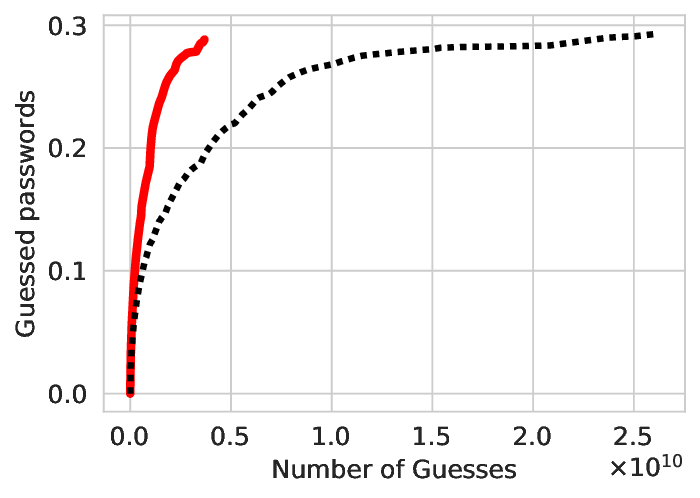}
		\caption{\textit{animoto} on \textit{MyHeritage}}\label{fig}
	\end{subfigure}\begin{subfigure}{.22\textwidth}
		\centering
		\includegraphics[trim = 0mm 0mm 0mm 0mm, clip, width=1\linewidth]{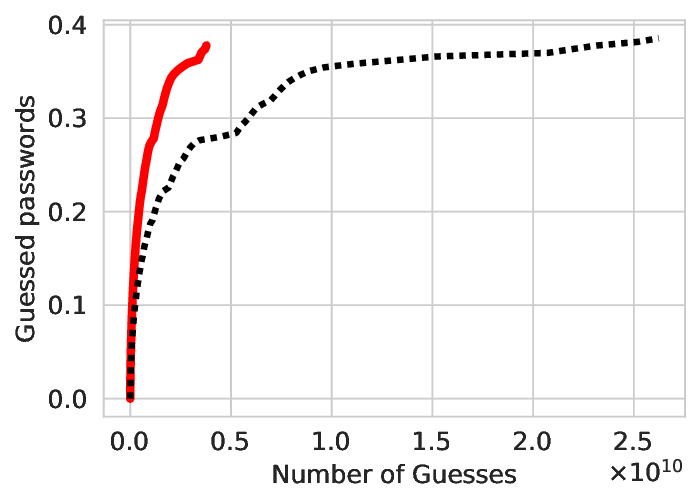}
		\caption{\textit{animoto} on \textit{RockYou}}\label{fig}
	\end{subfigure}\begin{subfigure}{.22\textwidth}
		\centering
		\includegraphics[trim = 0mm 0mm 0mm 0mm, clip, width=1\linewidth]{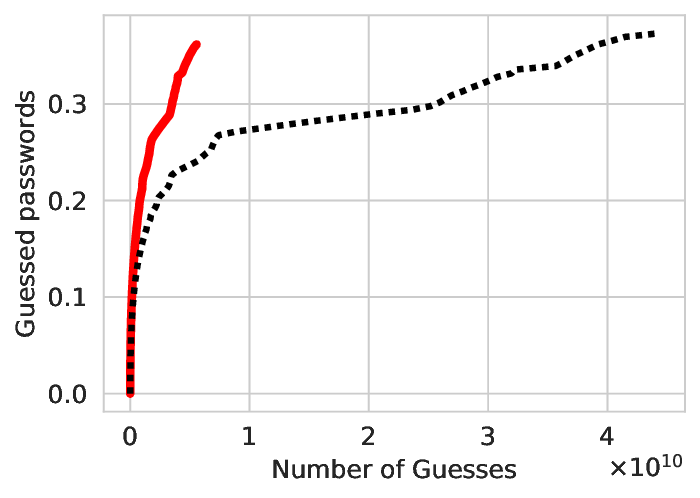}
		\caption{\textit{RockYou} on \textit{animoto}}\label{fig}
	\end{subfigure}
	\caption{Comparison between adaptive and classic mangling rules on four combination password leaks (dictionary/attacked-set) using the rules-set \textit{PasswordPro}. $\beta\myeq 0.5$ is used for the adaptive case.}
	\label{fig:ad_vs_cl}
\end{figure*}
As motivated in Section~\ref{sec:affinity}, each word in the dictionary interacts just with a limited number of mangling transformations that are conditionally defined by users' composition habits. While modern rules-sets can contain more than ten thousand entries, each \dw~$w$ will interact only with a small subset of \textbf{compatible rules}, say $R_w$. As stated before, optimized configurations compose over pairs of dictionaries and rule-sets that have been created to mutually support each other. This is achieved by implicitly maximizing the average cardinality of the compatible set of rules $R_w$ for each \dw~$w$ in the dictionary.
\par

In doing so, advanced attackers rely on domain knowledge and intuition to create optimized configurations. But, thanks to the explicit form of the compatibility function, it is possible to simulate their expertise. The intuition is that, given a \dw~$w$, we can infer the \textbf{compatible rules-set} $R_w$ (\ie the set of rules that interact well with $w$) according to the compatibility scores assigned by the neural approximation of~$\af$. More formally, given $\af$ for the rules-set $R$ and a \dw~$w$, we can determine the compatible rules-set for $w$ by \textit{thresholding} the compatibility values assigned by the neural network to the rules in $R$:
\begin{equation}
R_w\approx R_w^\beta=\{r\mid r \in R \wedge \af(w, r) > (1-\beta)\},
\label{ew:crs}
\end{equation}
where $\beta\in(0,1]$ is a threshold parameter whose effect will be discussed later.
\par
At this point, we simulate high-quality configuration attacks by ensuring \dws~does not interact with rules outside its compatible rules-set $R_w^\beta$. Algorithm~\ref{algo:amr} implements this strategy by following a word-major order in the generation of guesses. Every \dw~is limited to interact with the subset of compatible rules $R_w^\beta$ that is decided by the neural net. \textbf{Intuitively, this is equivalent to assigning and applying a dedicated (and possibly unique) rules-set to each word in the dictionary.} Note that, the selection of the compatible rules-set is performed at runtime, during the attack, and does not require any pre-computation. We call this novel guessing strategy \textbf{Adaptive Mangling Rules}, since the rule-set is continuously adapted during the attack to better assist the selected dictionary.
\par
\begin{figure}[t]
	\centering
	\includegraphics[trim = 0mm 0mm 0mm 0mm, clip, width=1\linewidth]{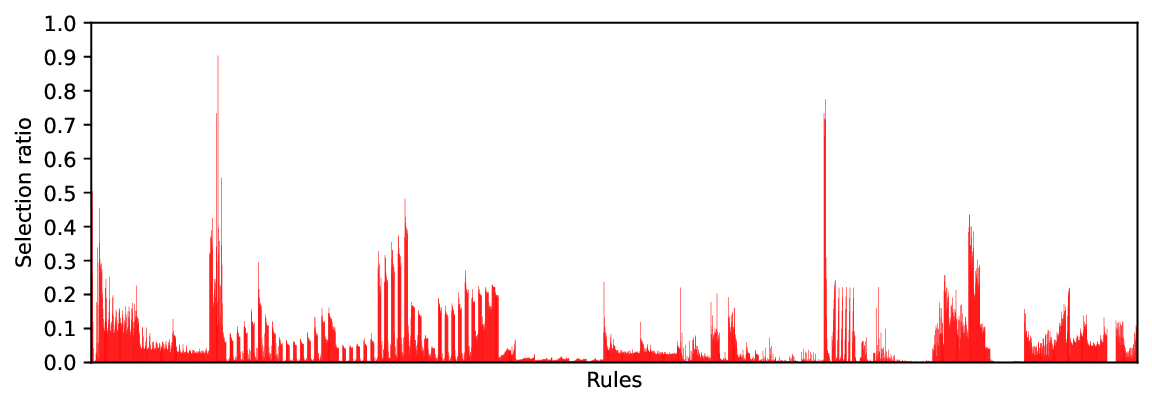}
	\caption{Selection frequencies of adaptive mangling rules for the 3120 rules of \textit{PasswordPro}.}
	\label{fig:hist}
\end{figure}

\begin{algorithm}[b]
	\KwData{dictonary $D$, rules-set $R$, budget $\beta$, neural net $\af_\RS$}
	\ForAll{$w\in D$}{
		$R_w^\beta = \{r|\af_\RS(w)_r >(1-\beta)\}$\;
		\ForAll{$r\in R_w^\beta$}{
			$g = r(w)$\;
			\textbf{issue} $g$\;
		}
	}
	\caption{Adaptive mangling rules attack.}
	\label{algo:amr}
\end{algorithm}
The efficacy of {\em adaptive} mangling rules over the standard attack is shown in Figure~\ref{fig:ad_vs_cl}, where multiple examples are reported. The adaptive mangling rules reduce the number of produced guesses while maintaining the hits count mostly unchanged.
 In our experiments, the adaptive approach induces compatible rules-sets that, on average, are an order of magnitude smaller than the complete rules-set. Typically, for $\beta\myeq0.5$, only $\sim10\%/15\%$ of the rules are conditionally applied to the \dws. Considering the percentage of guessed passwords for adaptive and non-adaptive attacks, this means that approximately $90\%$ of guesses are wasted during classic, unoptimized mangling rules attacks. Figure~\ref{fig:hist} further reports the distribution of selected rules during the adaptive attack of Figure~\ref{fig:ad_vs_cl_a}. It emphasizes how mangling rules heterogeneously interact with the underlying dictionary. Although very few rules interact well with all the words (\eg selection frequency is $>70\%$), most of the mangling rules participate only in rare events.
  \par
 
  Further empirical validation for the adaptive mangling rules will be given later in Section~\ref{sec:adam}.
\begin{figure*}[h]
	
	\centering
	\includegraphics[trim = 0mm 77mm 0mm 0mm, clip, width=.7\linewidth]{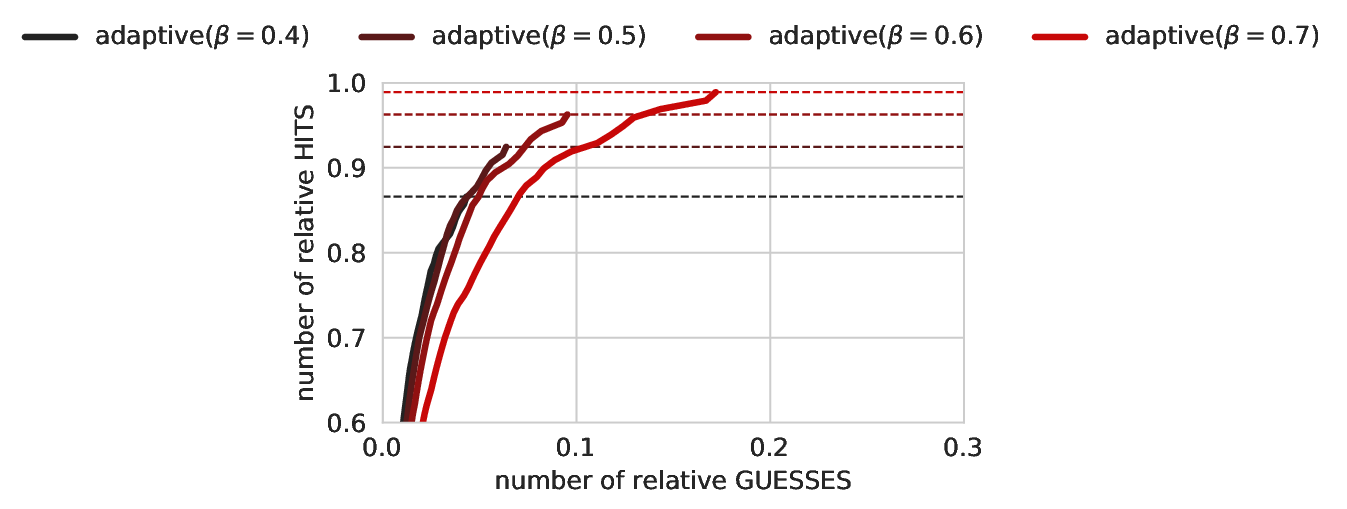}\\
	
	\begin{subfigure}{.23\textwidth}
		\centering
		\includegraphics[trim = 0mm 0mm 0mm 0mm, clip, width=1\linewidth]{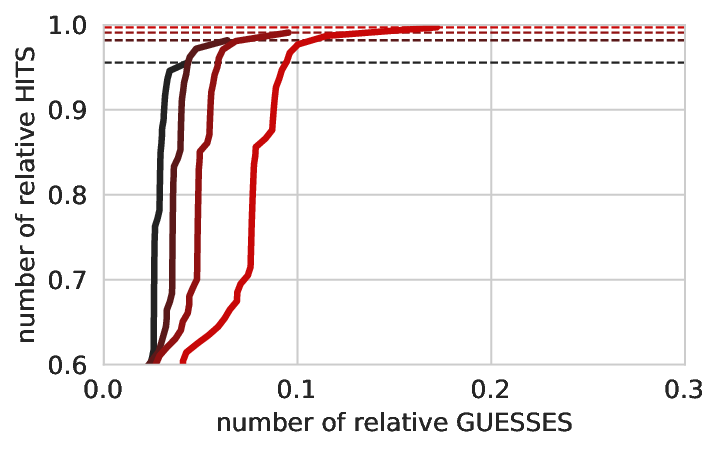}
		\caption{\textit{MyHeritage} on \textit{animoto}}\label{fig}
	\end{subfigure}\begin{subfigure}{.23\textwidth}
		\centering
		\includegraphics[trim = 0mm 0mm 0mm 0mm, clip, width=1\linewidth]{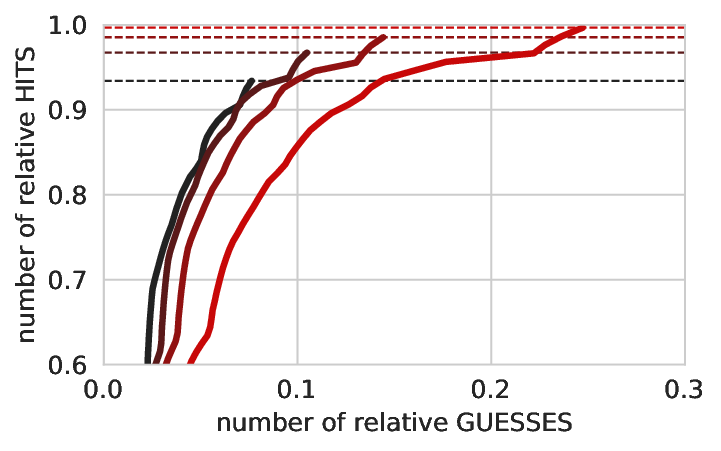}
		\caption{\textit{animoto} on \textit{MyHeritage}}\label{fig}
	\end{subfigure}\begin{subfigure}{.23\textwidth}
		\centering
		\includegraphics[trim = 0mm 0mm 0mm 0mm, clip, width=1\linewidth]{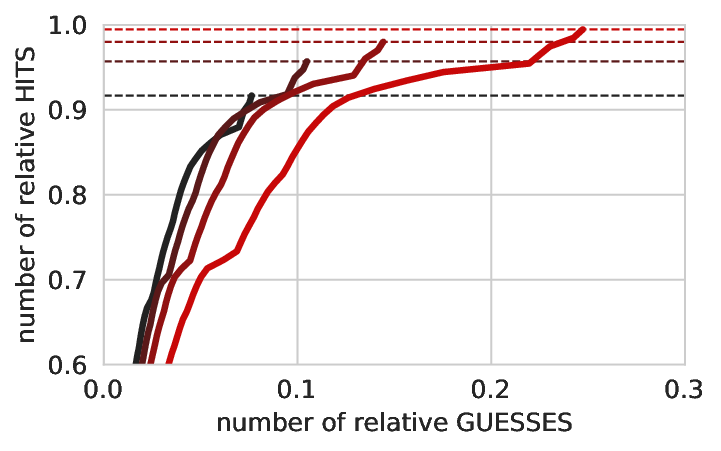}
		\caption{\textit{animoto} on \textit{RockYou}}\label{fig}
	\end{subfigure}\begin{subfigure}{.23\textwidth}
		\centering
		\includegraphics[trim = 0mm 0mm 0mm 0mm, clip, width=1\linewidth]{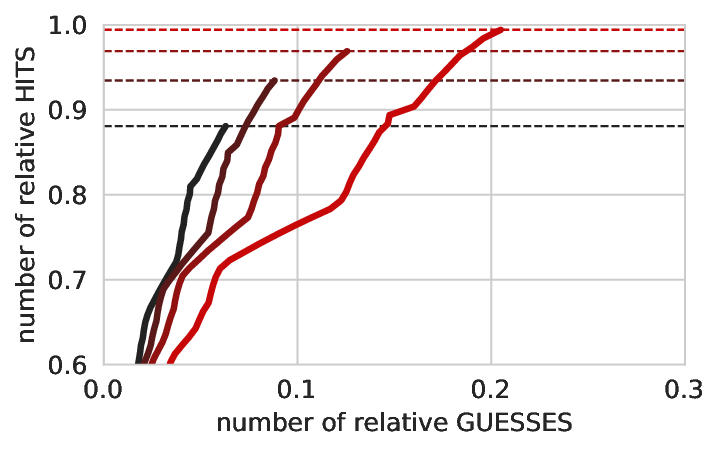}
		\caption{\textit{RockYou} on \textit{animoto}}\label{fig}
	\end{subfigure}
	
	\caption{Effect of the parameter $\beta$ on the guessing performance for four different combinations of password sets and \textit{PasswordPro} rules. Plots are normalized according to the results of the standard mangling rules attack (\ie $\beta=1$). For instance, $(x\myeq 0.1,\ y\myeq 0.95)$ means that we guessed $95\%$ of the password guessed with the standard mangling rules attack by performing $10\%$ of the guesses required from the latter.}
	\label{fig:beta_eff}
\end{figure*}
\paragraph{The Attack Budget}
Unlike standard dictionary attacks, whose effectiveness solely depends on the initial configuration, adaptive mangling rules can be controlled by an additional scalar parameter that we refer to as the \textbf{attack budget}~$\beta$. This parameter defines the threshold of compatibility that a rule must exceed to be included in the rules-set $R_w^\beta$ for a word~$w$. Indirectly, this value determines the average size of compatible rules-sets, and consequently, the total number of guesses performed during the attack. 
More precisely, low values of $\beta$ force compatible rule-sets to include only rules with high-compatibility. Those will produce only a limited number of guesses, inducing very precise attacks at the cost of missing possible hits (\ie high precision, low recall). Higher values of $\beta$ translate in a more permissive selection, where also rules with low-compatibility are included in the compatible set. Those will increase the number of produced guesses, inducing more exhaustive, yet more imprecise, attacks (\ie higher recall, lower precision). When $\beta$ reaches $1$, the adaptive mangling rules attack becomes a standard mangling rules attack, since all the rules are unconditionally included in the compatible rules-set. The effect of the budget parameter is better captured by the examples reported in Figure~\ref{fig:beta_eff}. Here, the performance of multiple values of $\beta$ is visualized and compared with the total hits and guesses performed by a standard mangling rules attack.
\par

The budget parameter $\beta$ can be used to model different types of adversaries. For instance, rational attackers~\cite{economics_pg} change their configuration in consideration of the practical cost of performing the attack. This parameter permit to easily describe those attackers and evaluate password security accordingly. For instance, using a low budget (\eg $\beta\myeq0.4$), we can model a greedy attacker who selects an attack configuration that maximizes guessing precision at the expense of the number of compromised accounts (a rational behavior in case of an expensive hash function).
\par

Seeking a more pragmatic interpretation, the budget parameter is implicitly equivalent to \textit{early-stopping}\footnote{The attack stops before the guesses are terminated.} (\ie Eq.~\ref{eq:bsr}), where single guesses are sorted in optimal order \ie guesses are exhaustively generated before the attack, and indirectly sorted by decreasing probability/compatibility. 
\par

The optimal value of $\beta$ depends on the rules-set. In our tests, we found these optimal values to be $0.6,\ 0.8$ and $0.8$ for \textit{PassowordPro}, \textit{generated} and \textit{generated2}, respectively. Hereafter, we use these setups, unless otherwise specified.
\paragraph{Computational cost}
One of the core advantages of dictionary attacks over more sophisticated approaches \cite{FLA, PCFG, MM} is their speed. For mangling rules attacks, generating guesses has almost a negligible impact. Despite being consistently more complex in their mechanisms, adaptive mangling rules do not tend to change this feature.
\par

In Algorithm~\ref{algo:amr}, the only additional operation over the standard mangling rules attack is the selection of compatible rules for each \dw~via the trained neural net. As discussed in Section~\ref{sec:learning_aff}, this operation requires just a single network inference to be computed; that is, with a single inference, we obtain a compatibility score for each element in $\{w\}\times R$. Furthermore, inference for multiple consecutive words can be trivially batched and computed in parallel, further reducing the computation's impact.\par

Table~\ref{tab:csa} reports the number of compatibility values that different neural networks can compute per second. 
\begin{table}[b]
	\centering
	\caption{Number of compatible scores computed per second (c/s) for different networks. Values computed on a single NVIDIA V100 GPU.}
	\label{tab:csa}
	\resizebox{.8\columnwidth}{!}{%
		\label{table:perf_data}
		\begin{tabular}{cccccc}
			\toprule
			\makecell{ \textbf{generated2} \\ \textit{(large)}} & \makecell{ \textbf{generated} \\ \textit{(large)}} & \makecell{ \textbf{PasswordPro} \\ \textit{(large)}} \\ \midrule
			$130.550.403$ c/s & $89.049.382$ c/s & $31.836.734$ c/s \\
			\bottomrule
		\end{tabular}
	}
\end{table}
In the table, we used our largest networks without any form of optimization. Nevertheless, the overhead over the plain mangling rules attack is minimal (see Appendix~\ref{app:bench}). 
Additionally, similar to standard dictionary attacks, adaptive mangling rules attacks are inherently parallel and, therefore, distributed and scalable.
\par

%
	\section{Dynamic Dictionary attacks}
	\label{sec:dynamic}
	This section introduces the second and last component of our password model---a dynamic mechanism that systematically adapts the guessing configuration to the unknown attacked-set. In Section~\ref{sec:dyn_dict}, we introduce the \textbf{Dynamic Dictionary Augmentation} technique. Next, in Section~\ref{sec:dyn_bud}, we introduce the concept of a \textbf{Dynamic Budgets}.
\paragraph{Motivation:}
As widely documented~\cite{the_science_of_guessing, emeprical_anal, muni, imprl}, password composition habits slightly change from sub-population to sub-population. Although passwords tend to follow the same general distribution, credentials created under different environments exhibit unique biases. Users within the same group usually choose passwords related to each other, influenced mostly by environmental factors or the underlying applicative layer. Major factors, for example, are users' mother tongue~\cite{emeprical_anal}, community interests~\cite{sem_biases} and, imposed password composition policies~\cite{composition_policy}. These have a significant impact on defining the final password distribution, and, consequently, the \textit{guessability} of the passwords~\cite{guess}. The same factors that shape a password distribution are generally available to the attackers who can collect and use them to drastically improve the configuration of their guessing attacks.
Unfortunately, current automatic guessing techniques fail to describe this natural adversarial behavior \cite{zxcvb, guess, muni, measuring, reasoning}. Those methods are based on static configurations that apply the same guessing strategy to each attacked-set of passwords, mostly ignoring trivial information that can be either \textit{a priori} collected or distilled from the running attack. 
In this section, we discuss suitable modifications of the mangling-rules framework to describe a more realistic guessing strategy. In particular, avoiding the necessity of any prior knowledge over the attacked-set, we rely on the concept of \textbf{dynamic attack}~\cite{imprl}. Here, a dynamic attacker is an adversary who changes his guessing strategy according to the attack's success rate. Successful guesses are used to select future attempts with the goal of exploiting the non-i.i.d. of passwords originated from the same environment. 
In other words, dynamic password guessing attacks automatically collect information on the target password distribution and use it to forge unique guessing configurations for the same set during the attack. 
Similarly, this general guessing approach can be easily linked to the optimal guessing strategy harnessed from human experts in \cite{measuring}, where mangling rules were manually created at execution time based on the initially guessed passwords.
\subsection{Dynamic Dictionary Augmentation}
\label{sec:dyn_dict}
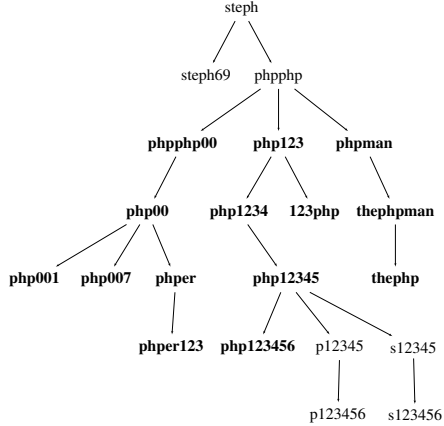
\begin{figure}[t]
		\centering
		\resizebox{.7\columnwidth}{!}{%
		\begin{tikzpicture}[>=latex',line join=bevel, every node/.style={scale=1.7}]
	\node (1) at (249.0bp,522.0bp) [draw,draw=none,opacity=1] {steph};
	\node (2) at (211.0bp,450.0bp) [draw,draw=none,opacity=1] {steph69};
	\node (10) at (288.0bp,450.0bp) [draw,draw=none] {phpphp};
	\node (11) at (186.0bp,378.0bp) [draw,draw=none] {\textbf{phpphp00}};
	\node (12) at (288.0bp,378.0bp) [draw,draw=none] {\textbf{php123}};
	\node (13) at (379.0bp,378.0bp) [draw,draw=none] {\textbf{phpman}};
	\node (60) at (149.0bp,306.0bp) [draw,draw=none] {\textbf{php00}};
	\node (16) at (246.0bp,306.0bp) [draw,draw=none] {\textbf{php1234}};
	\node (17) at (326.0bp,306.0bp) [draw,draw=none] {\textbf{123php}};
	\node (14) at (412.0bp,306.0bp) [draw,draw=none] {\textbf{thephpman}};
	\node (15) at (412.0bp,234.0bp) [draw,draw=none] {\textbf{thephp}};
	\node (18) at (296.0bp,234.0bp) [draw,draw=none] {\textbf{php12345}};
	\node (19) at (266.0bp,162.0bp) [draw,draw=none] {\textbf{php123456}};
	\node (20) at (353.0bp,162.0bp) [draw,draw=none] {p12345};
	\node (21) at (431.0bp,162.0bp) [draw,draw=none] {s12345};
	\node (24) at (351.0bp,90.0bp) [draw,draw=none] {p123456};
	\node (22) at (433.0bp,90.0bp) [draw,draw=none] {s123456};
	\node (61) at (29.0bp,234.0bp) [draw,draw=none] {\textbf{php001}};
	\node (62) at (105.0bp,234.0bp) [draw,draw=none] {\textbf{php007}};
	\node (63) at (179.0bp,234.0bp) [draw,draw=none] {\textbf{phper}};
	\node (64) at (174.0bp,162.0bp) [draw,draw=none] {\textbf{phper123}};
	\draw [->, opacity=1] (1) ..controls (235.24bp,495.64bp) and (229.94bp,485.89bp) .. (2);
	\draw [->,opacity=1] (1) ..controls (263.13bp,495.64bp) and (268.56bp,485.89bp) .. (10);
	\draw [->] (10) ..controls (249.96bp,422.89bp) and (233.74bp,411.76bp) .. (11);
	\draw [->] (10) ..controls (288.0bp,423.98bp) and (288.0bp,414.71bp) .. (12);
	\draw [->] (10) ..controls (321.96bp,422.88bp) and (336.06bp,412.03bp) .. (13);
	\draw [->] (11) ..controls (172.6bp,351.64bp) and (167.44bp,341.89bp) .. (60);
	\draw [->] (12) ..controls (272.74bp,351.56bp) and (266.81bp,341.69bp) .. (16);
	\draw [->] (12) ..controls (301.76bp,351.64bp) and (307.06bp,341.89bp) .. (17);
	\draw [->] (13) ..controls (390.91bp,351.73bp) and (395.45bp,342.1bp) .. (14);
	\draw [->] (14) ..controls (412.0bp,279.98bp) and (412.0bp,270.71bp) .. (15);
	\draw [->] (16) ..controls (264.23bp,279.47bp) and (271.37bp,269.48bp) .. (18);
	\draw [->] (18) ..controls (285.21bp,207.81bp) and (281.13bp,198.3bp) .. (19);
	\draw [->] (18) ..controls (316.92bp,207.3bp) and (325.26bp,197.07bp) .. (20);
	\draw [->] (18) ..controls (348.52bp,205.77bp) and (373.26bp,192.94bp) .. (21);
	\draw [->] (20) ..controls (352.29bp,135.98bp) and (352.02bp,126.71bp) .. (24);
	\draw [->] (21) ..controls (431.71bp,135.98bp) and (431.98bp,126.71bp) .. (22);
	\draw [->] (60) ..controls (105.64bp,279.7bp) and (84.94bp,267.63bp) .. (61);
	\draw [->] (60) ..controls (133.01bp,279.56bp) and (126.8bp,269.69bp) .. (62);
	\draw [->] (60) ..controls (159.79bp,279.81bp) and (163.87bp,270.3bp) .. (63);
	\draw [->] (63) ..controls (177.21bp,207.98bp) and (176.55bp,198.71bp) .. (64);
		\end{tikzpicture}
	}

	\caption{Example of small \textit{hits-tree} induced by the dynamic attack performed on the \textit{phpBB} leak. In the tree, every vertex is a guessed password; an edge between two nodes indicates that the child password has been guessed by applying a mangling rule to the parent password.}
	\label{fig:hit_tree}
	\end{figure}

In \cite{imprl}, dynamic adaptation of the guessing strategy is obtained from password latent space manipulations of deep generative models. A similar effect is reproduced within our mangling rules approach by relying on a consistently simpler, yet effective, solution based on \textit{hits-recycling}. That is, every time we guess a new password by applying a mangling rule over a dictionary word, we insert the guessed password in the dictionary at runtime. In practice, \textbf{we dynamically augment the dictionary during the attack using the guessed passwords}.\footnote{Although we have not found any direct reference to the \textit{hits-recycling} technique in the literature, it is likely well known and routinely deployed by professionals.} In the process, every new hit is directly reconsidered and syntactically extended through mangling rules. This recursive method brings about massive chains/trees of hits that can extend for thousands of levels.\footnote{I.e., a forest, where the root of each tree is a word from the original dictionary.} Figure \ref{fig:hit_tree} depicts an extremely small subtree (\TT{\textit{hits-tree}}) obtained by attacking the password leak \textit{phpBB}. The tree starts when the word $\TT{steph}$ is mangled, incidentally producing the word $\TT{phpphp}$. Since the latter lies in a dense zone of the attacked set (\ie it is a common users' practice to insert the name of the website or related strings in their password), it induces multiple hits and causes the attack to focus in that specific zone of the key-space. The focus of the attack grows exponentially hit after hit and automatically stops only when no more passwords are matched. Eventually, this process makes it possible to guess passwords that would be missed with the static approach. For instance, in Figure~\ref{fig:hit_tree}, all the nodes in bold are passwords matched by the \textit{dynamic} attack but missed by the \textit{static} one (\ie standard dictionary attack) under the same configuration.
\par
\begin{figure*}[t]
	\centering
	\begin{subfigure}{0.22\textwidth}
		\includegraphics[trim = 30mm 88mm 15mm 0mm, clip, width=.9\linewidth]{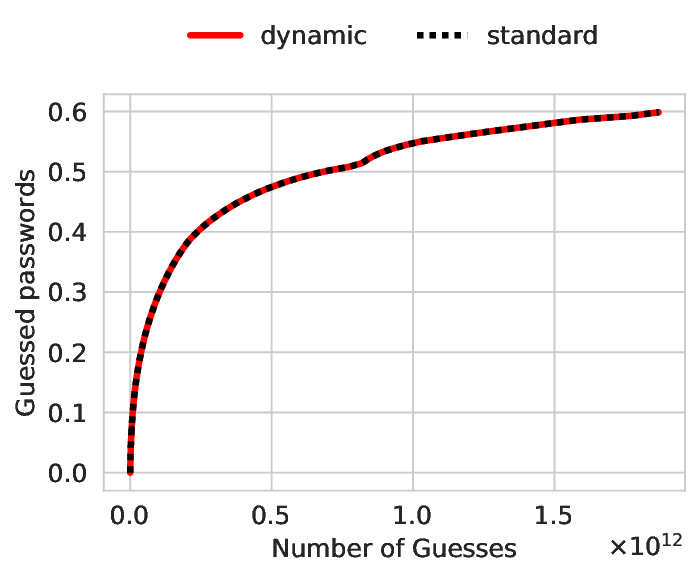}
	\end{subfigure} 
	\par
	\resizebox{1\textwidth}{!}{
		
		\begin{tabular}{ccccc}
			\begin{subfigure}{0.22\textwidth}
				\centering
				\includegraphics[trim = 0mm 0mm 0mm 0mm, clip, width=1\linewidth]{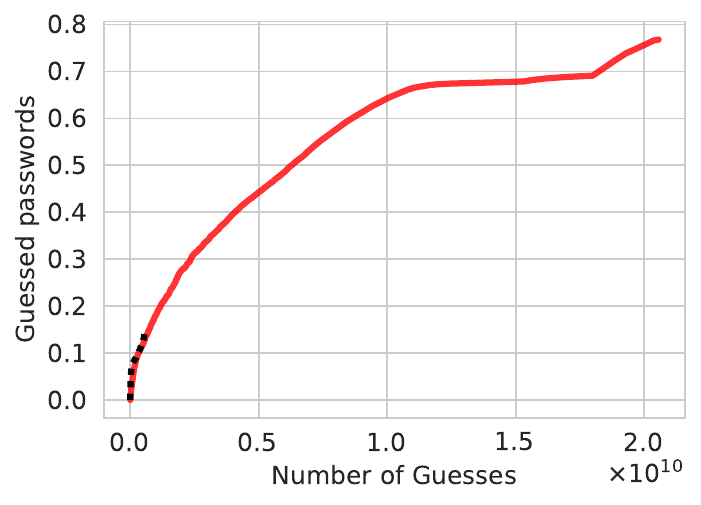}
				\caption{\textit{phpBB} on \textit{animoto}}\label{tbl:dyn_vs_static_password_pr_a}
			\end{subfigure} &
			\begin{subfigure}{0.22\textwidth}
				\centering
				\includegraphics[trim = 0mm 0mm 0mm 0mm, clip, width=1\linewidth]{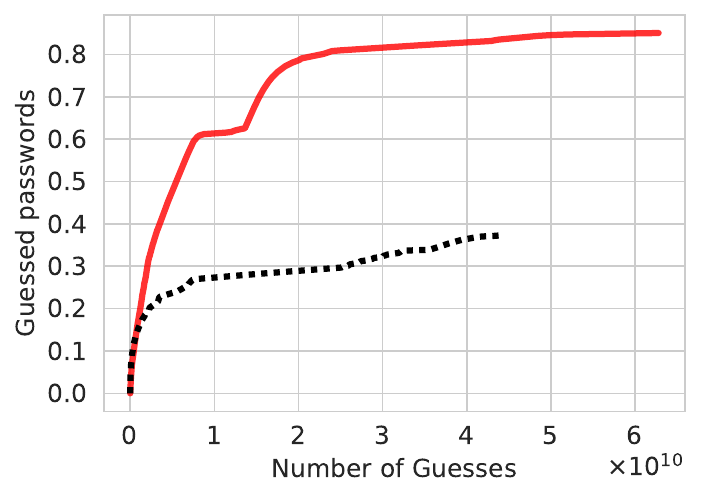}
				\caption{\textit{RockYou} on \textit{animoto}}\label{tbl:dyn_vs_static_password_pr_b}
			\end{subfigure}
			&
			\begin{subfigure}{0.22\textwidth}
				\centering
				\includegraphics[trim = 0mm 0mm 0mm 0mm, clip, width=1\linewidth]{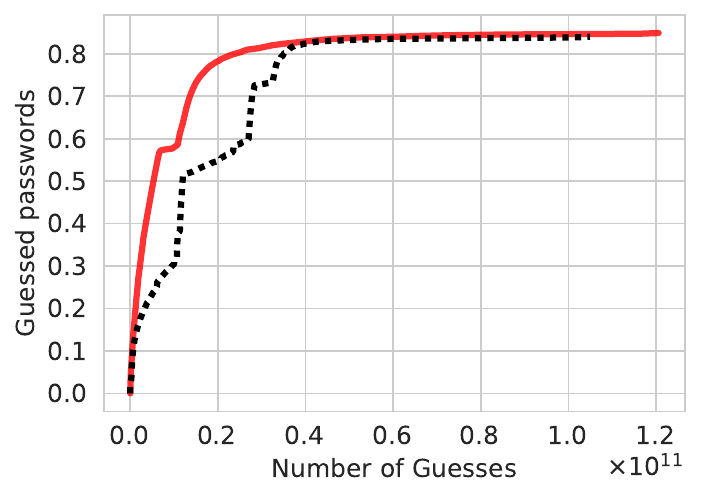}
				\caption{\textit{MyHeritage} on \textit{animoto}}\label{tbl:dyn_vs_static_password_pr_c}
			\end{subfigure}	 &
			\begin{subfigure}{0.22\textwidth}
				\centering
				\includegraphics[trim = 0mm 0mm 0mm 0mm, clip, width=1\linewidth]{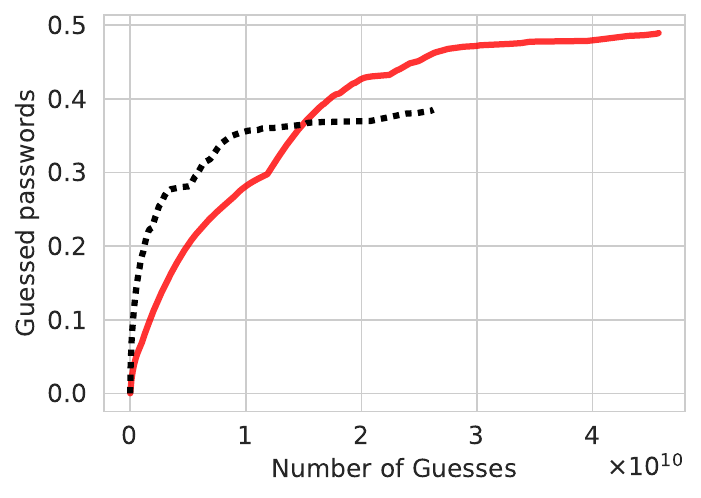}
				\caption{\textit{animoto} on \textit{RockYou}}\label{tbl:dyn_vs_static_password_pr_d}
			\end{subfigure} &
			\begin{subfigure}{0.22\textwidth}
				\centering
				\includegraphics[trim = 0mm 0mm 0mm 0mm, clip, width=1\linewidth]{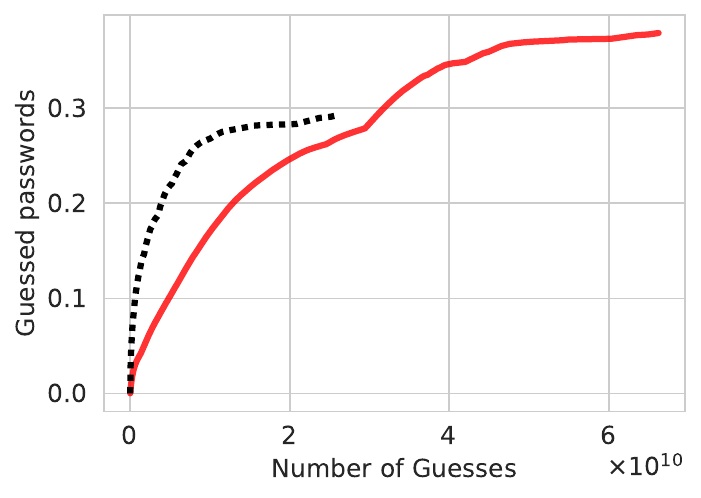}
				\caption{\textit{animoto} on \textit{MyHeritage}}\label{tbl:dyn_vs_static_password_pr_e}
			\end{subfigure} 
		\end{tabular}
	}
	\caption{Performance comparison between dynamic and standard (static) attack for five different setups of dictionary/attacked-set. The rules set \textit{PasswordPro} in {non-adaptive mode} is used in all the reported attacks. The $5$ setups have been handpicked to fully represent the possible effects of the dynamic dictionary augmentation.}
	\label{tbl:dyn_vs_static_password_pro}
\end{figure*}
Figure \ref{tbl:dyn_vs_static_password_pro} compares the guessing performance of the dynamic attack against the static version on a few examples for the \textit{PasswordPro} rules-set. The plots show that the dynamic augmentation of the dictionary has a very heterogeneous effect on the guessing attacks. In the case of Figure~\ref{tbl:dyn_vs_static_password_pr_a}, the dynamic attack produces a substantial increment in the number of guesses as well as in the number of hits \ie from $\sim15\%$ to $\sim80\%$ recovered passwords. Arguably, such a gap is due to the minimal size of the original dictionary \textit{phpBB}. In the attack of Figure \ref{tbl:dyn_vs_static_password_pr_b}, instead, a similar improvement is achieved by requiring only a small number of guesses. On the other hand, in the attack depicted in Figure \ref{tbl:dyn_vs_static_password_pr_c}, the dynamic augmentation has a limited effect on the final hits number. However, it increases the attack precision in the initial phase. Conversely, attacks in Figures \ref{tbl:dyn_vs_static_password_pr_d} and \ref{tbl:dyn_vs_static_password_pr_e} show a decreased precision in the initial phase of the attack, but that is compensated later by the dynamic approach. 
\par

\begin{figure}[h]
	\centering
	\includegraphics[trim = 0mm 93mm 0mm 0mm, clip, width=.6\columnwidth]{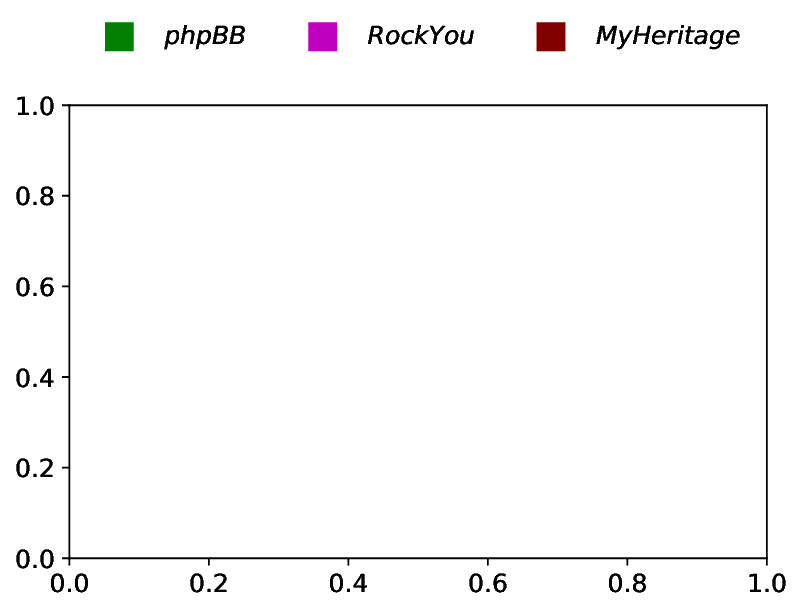}\\
	
	\begin{subfigure}{.25\textwidth}
		\centering
		\includegraphics[trim = 0mm 0mm 0mm 0mm, clip, width=.9\linewidth]{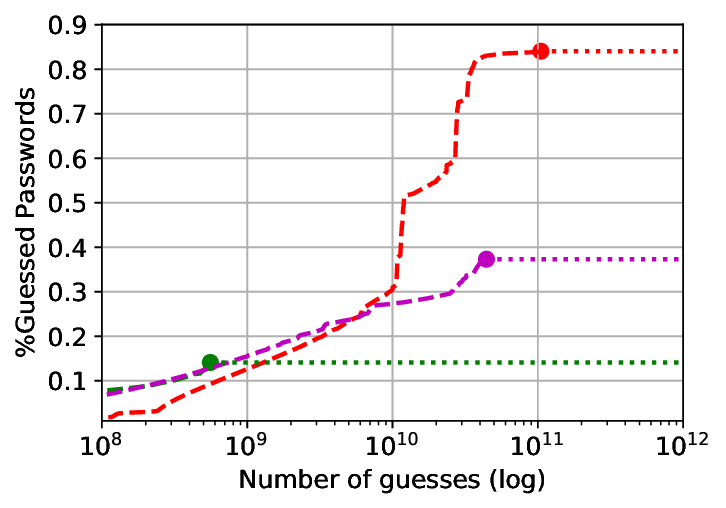}
		\caption{standard attack}\label{fig:dyn_vs_stat_hetero_a}
	\end{subfigure}\begin{subfigure}{.25\textwidth}
		\centering
		\includegraphics[trim = 0mm 0mm 0mm 0mm, clip, width=.9\linewidth]{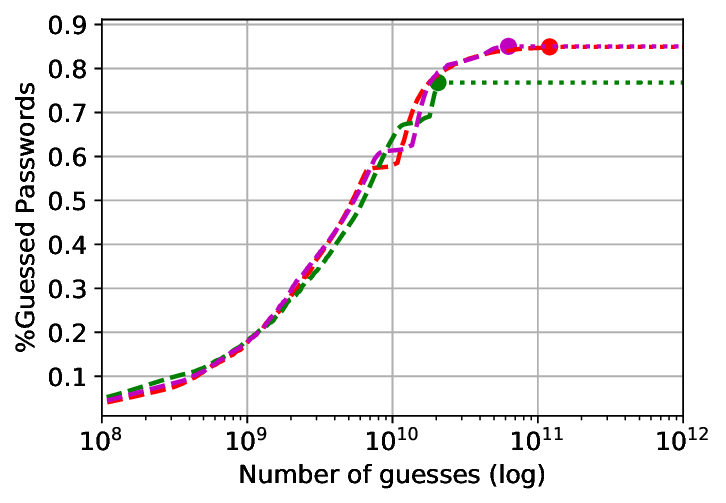}
		\caption{dynamic attack}\label{fig:dyn_vs_stat_hetero_b}
	\end{subfigure}
\caption{Guessing attacks performed on the \textit{animoto} leak using three different dictionaries. The panel on the left reports the guessing curves for the static setup. The panel on the right reports those for the dynamic setup. The x-axis is logarithmic.}
\label{fig:dyn_vs_stat_hetero}
\end{figure}

Another interesting property of the dynamic augmentation is that it makes the guessing attack consistently less sensitive to the choice of the input dictionary. Indeed, in contrast with the static approach, different choices of the initial dictionary tend to produce very homogeneous results in the dynamic approach. This behavior is captured in Figure \ref{fig:dyn_vs_stat_hetero}, where results, obtained by varying three input dictionaries, are compared between static and dynamic attack. The standard attacks (Figure \ref{fig:dyn_vs_stat_hetero_a}) result in very different outcomes; for instance, using \textit{phpBB} we match $15\%$ of the attacked-set, whereas we match more than $80\%$ with \textit{MyHeritage}. These differences in performance are leveled out by the dynamic augmentation of the dictionary (Figure \ref{fig:dyn_vs_stat_hetero_b}); all the dynamic attacks recover $\sim80\%$ of the attacked-set. Intuitively, dynamic augmentation remedies deficiencies in the initial configuration of the dictionary, promoting its completeness. These claims will find further support in Section~\ref{sec:adam}.
\subsection{Dynamic budgets}
\label{sec:dyn_bud}
Adaptive mangling rules (Section~\ref{sec:amra}) demonstrated that it is possible to consistently improve the precision of the guessing attack by promoting compatibility among rules-set and dictionary (\ie simulating high-quality configurations at runtime). This approach assumes that the compatibility function modeled before the attack is sufficiently general to simulate good configurations for each possible attacked-set. 
 However, as motivated in the introduction of Section~\ref{sec:dynamic}, every attacked set of passwords present peculiar biases and, therefore, different compatibility relations among rules and dictionary-words.\\
 To reduce the effect of this dependence, we introduce an additional dynamic approach supporting the adaptive mangling rules framework. Rather than modifying the neural network at runtime (which is neither a practical nor a reliable solution), we alter the selection process of compatible rules by acting on the budget parameter $\beta$.
 \par
\begin{algorithm}[b]
	\SetAlgoLined
	\KwData{\small dictonary $D$, rules-set $R$, attacked-set $X$, budget $\beta$}
	\ForAll{$w\in D$}{
		$R_w^\beta = \{r| \af_\RS(w)_r>(1-B_i)\}$\; 
		\ForAll{$r\in R_w^\beta$}{
			$g = r(w)$\;
			\If{$g \in X$}{
				$X = X - \{g\}$\;
				$B_r = B_r + \Delta$\;
			}
		$B = B\cdot \frac{\sum^{|B|}\beta}{\sum^{|B|}B}$\;
		}
	}
	\caption{Adaptive rules with Dynamic budget}
	\label{alg:dyn_b}
\end{algorithm}

Algorithm~\ref{alg:dyn_b} details our solution. Here, rather than having a global parameter $\beta$ for all the rules of the rules-set $R$, we have a budget vector $B$ that assigns a dedicated budget value to each rule in $\RS$ (\ie $B\in(0,1]^{|\RS|}$). Initially, all the budget values in $B$ are initialized to the same value $\beta$ (\ie $\forall_{r\in \RS}\ B_r\myeq \beta$) given as an input parameter. During the attack, the elements of $B$ are individually increased and decreased to better describe the attacked set of passwords. Within this context, increasing the budget $B_r$ of a rule $r$ means reducing the compatibility threshold needed to include $r$ in the compatible rules-set of a \dw~$w$, and, consequently, making $r$ more popular during the attack. On the other hand, by decreasing $B_r$, we reduce the chances of selection for $r$; $r$ is selected only in case of high-compatibility words.\par

In the algorithm, we increase the budget~$B_r$ when the rule~$r$ produces a hit. The added increment is a small value $\Delta$ that scales inversely with the number of guesses produced. At the end of the internal loop, the vector~$B$ is then normalized; \ie we scale the values in~$B$ so that $\sum_{r}^{\RS}B_r=\sum_{i}^{|\RS|}\beta$. Normalizing~$B$ has two aims. \textbf{(1)} It reduces the budgets for non-hitting rules (the mass we add to the budget of rule $r$ is subtracted from all other budgets). \textbf{(2)} It maintains the total budget of the attack (\ie $\sum_{i}^{|\RS|}\beta$) unchanged so that dynamic and static budget leads to almost the same number of guesses during the attack for a given $\beta$. 
Furthermore, we impose a maximum and a minimum bound on the increments or decrements of~$B$. This is to prevent values of zero (rule always excluded) or equal/higher than one (rule always included).
\par

As for the dynamic dictionary augmentation, the dynamic budget has always a positive, but, heterogeneous, effect on the guessing performance. Mostly, the number of hits increases or remains unaffected. Among the proposed techniques, this is the one with the mildest effect. Yet, this will be particularly useful when combined with dynamic dictionary augmentation in the next section. Appendix~\ref{app:dynbud} better explicates the improvement induced from the dynamic budgets.
	
	%
	%
%
\begin{figure*}[h!]
	\centering
	\begin{subfigure}{0.4\textwidth}
		\centering
		\includegraphics[trim = 6mm 90mm 5mm 0mm, clip, width=.65\linewidth]{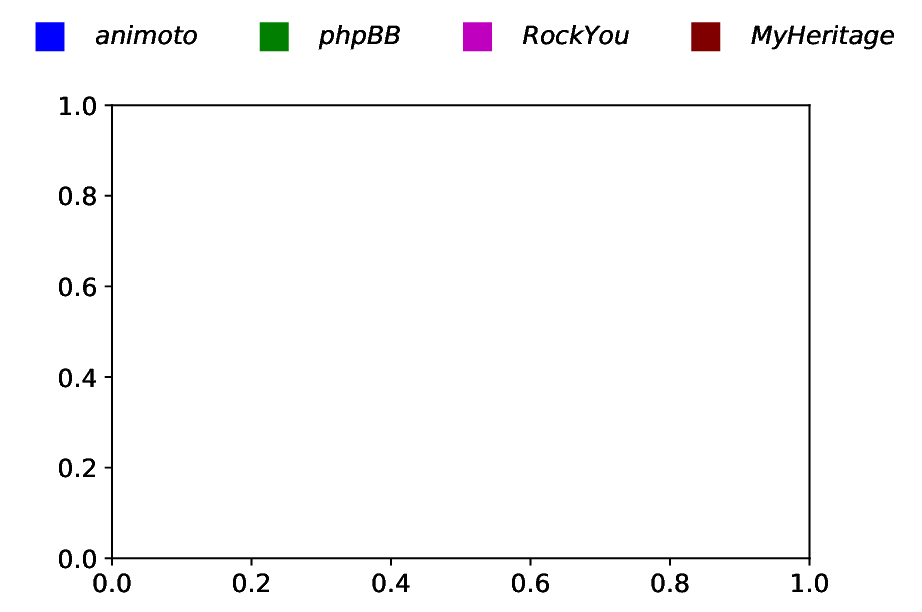}\includegraphics[trim = 35mm 91mm 15mm 0mm, clip, width=.45\linewidth]{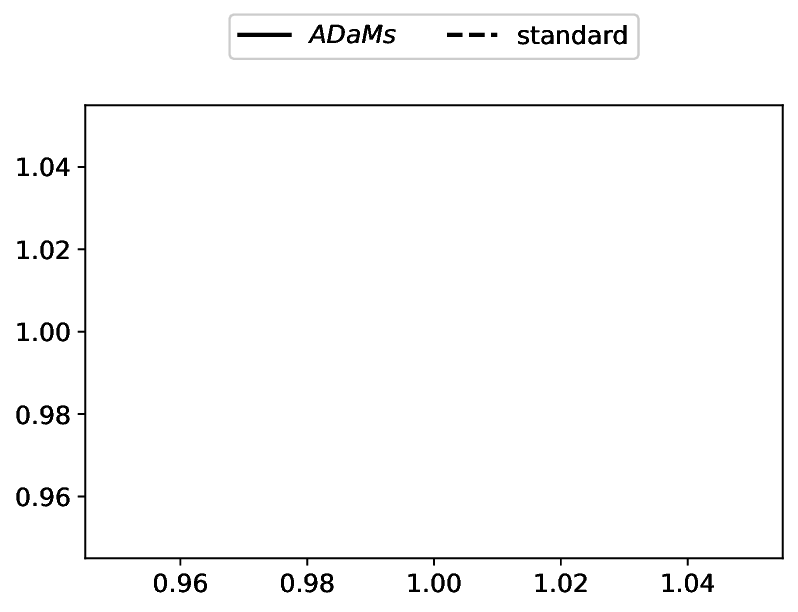}
		
	\end{subfigure} 
	\par
	\resizebox{1\textwidth}{!}{
		\begin{tabular}{cccc}
			\begin{subfigure}{0.25\textwidth}
				\centering
				\includegraphics[trim = 0mm 0mm 0mm 0mm, clip, width=1\linewidth]{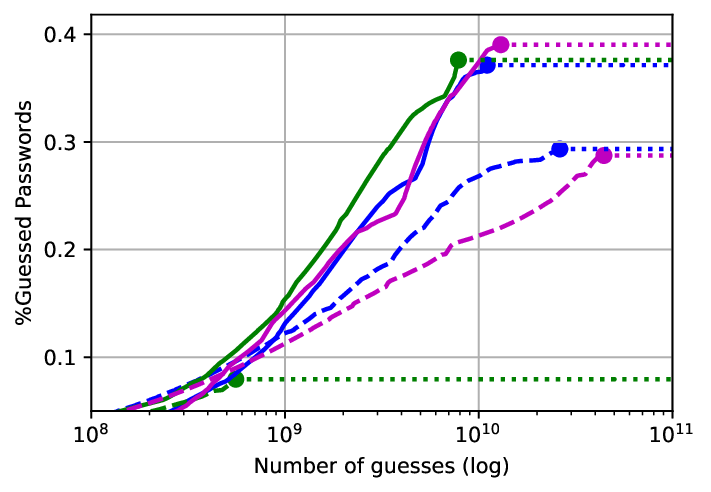}
				\vspace{-0.7cm}
				\caption{\textit{MyHeritage} with \textit{PasswordPro}}\label{fig:rmp}
			\end{subfigure} &
		\begin{subfigure}{0.25\textwidth}
			\centering
			\includegraphics[trim = 0mm 0mm 0mm 0mm, clip, width=1\linewidth]{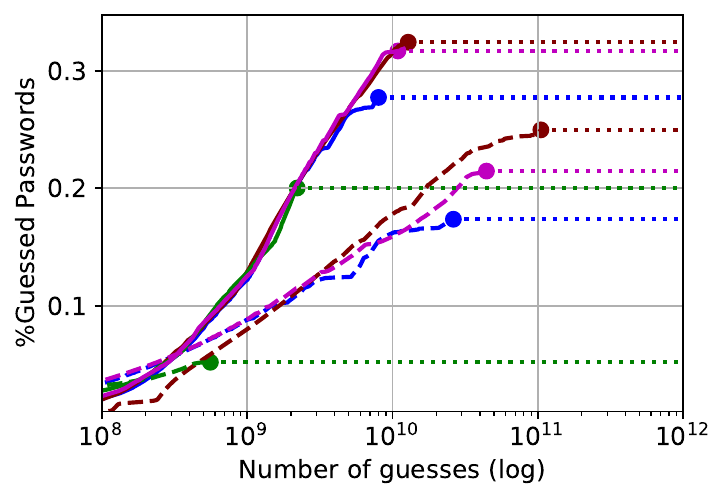}
			\vspace{-0.7cm}
			\caption{\textit{zooks} with \textit{PasswordPro}}\label{fig:rzp}
		\end{subfigure}&
	\begin{subfigure}{0.25\textwidth}
		\centering
		\includegraphics[trim = 0mm 0mm 0mm 0mm, clip, width=1\linewidth]{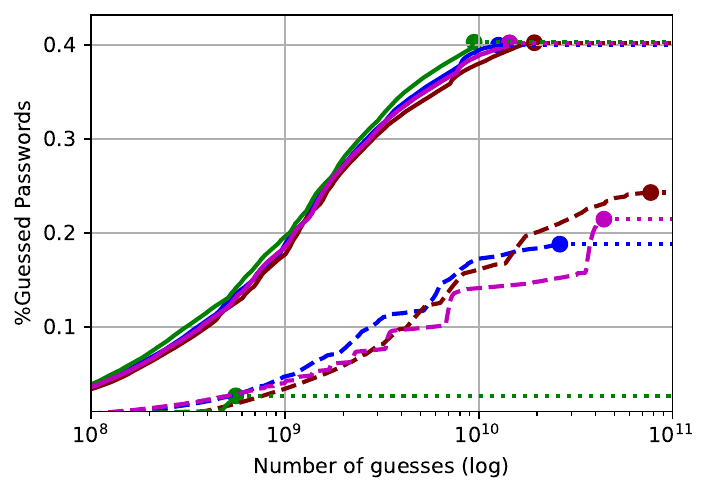}
		\vspace{-0.7cm}
		\caption{\textit{youku} with \textit{PasswordPro}}\label{fig:ryp}
	\end{subfigure}&
	\begin{subfigure}{0.25\textwidth}
	\centering
	\includegraphics[trim = 0mm 0mm 0mm 0mm, clip, width=1\linewidth]{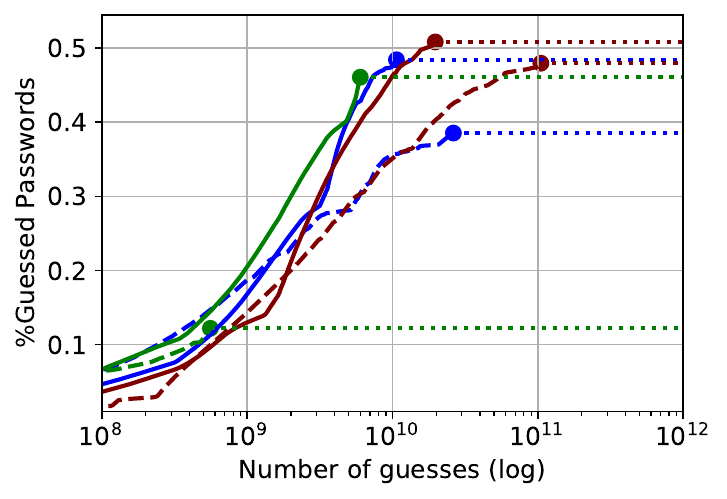}
	\vspace{-0.7cm}
	\caption{\textit{RockYou} with \textit{PasswordPro}}\label{fig:rrp}
\end{subfigure}\\
	\begin{subfigure}{0.25\textwidth}
		\centering
		\includegraphics[trim = 0mm 0mm 0mm 0mm, clip, width=1\linewidth]{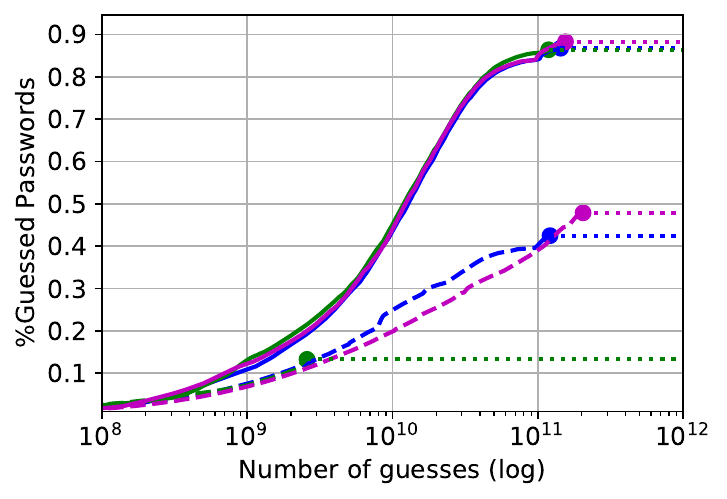}
		\vspace{-0.7cm}
		\caption{\textit{MyHeritage} with \textit{generated}}\label{fig:rmg}
	\end{subfigure} &
	\begin{subfigure}{0.25\textwidth}
		\centering
		\includegraphics[trim = 0mm 0mm 0mm 0mm, clip, width=1\linewidth]{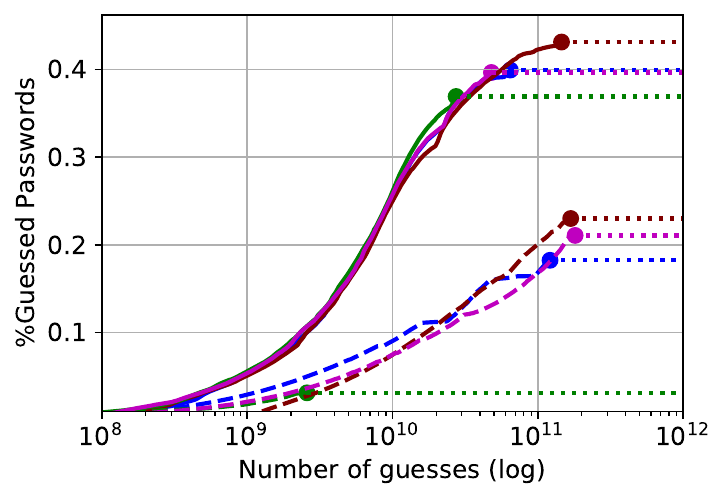}
		\vspace{-0.7cm}
		\caption{\textit{zooks} with \textit{generated}}\label{fig:rzg}
	\end{subfigure}&
	\begin{subfigure}{0.25\textwidth}
		\centering
		\includegraphics[trim = 0mm 0mm 0mm 0mm, clip, width=1\linewidth]{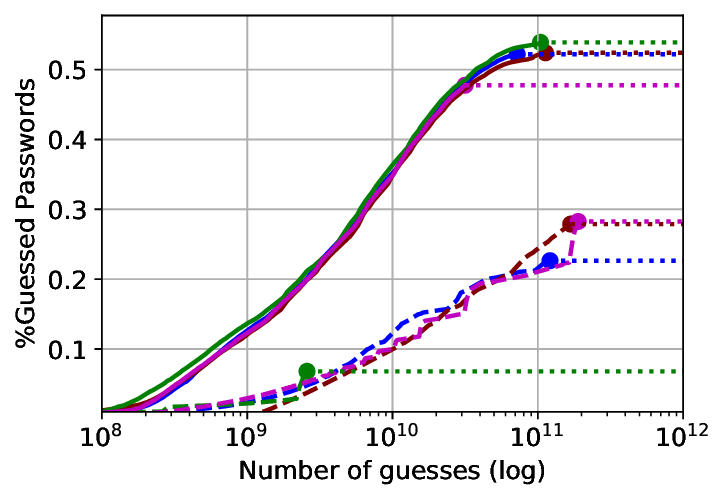}
		\vspace{-0.7cm}
		\caption{\textit{youku} with \textit{generated}}\label{fig:ryg}
	\end{subfigure}&
	\begin{subfigure}{0.25\textwidth}
		\centering
		\includegraphics[trim = 0mm 0mm 0mm 0mm, clip, width=1\linewidth]{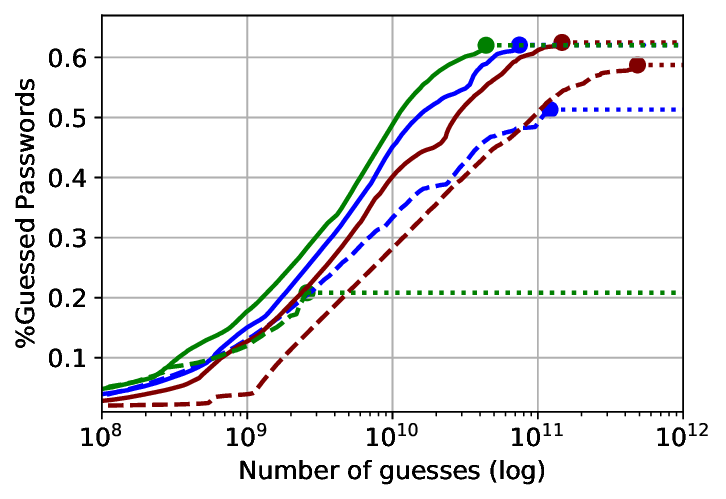}
		\vspace{-0.7cm}
		\caption{\textit{RockYou} with \textit{generated}}\label{fig:rrg}
	\end{subfigure}\\

			\begin{subfigure}{0.25\textwidth}
	\centering
	\includegraphics[trim = 0mm 0mm 0mm 0mm, clip, width=1\linewidth]{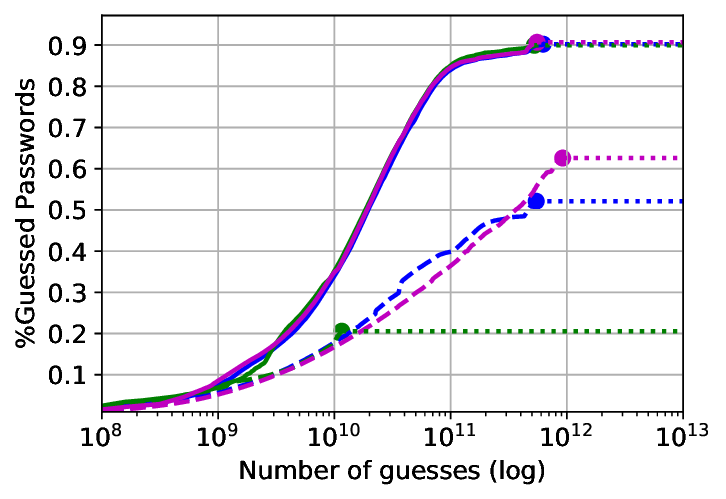}
	\vspace{-0.7cm}
	\caption{\textit{MyHeritage} with \textit{generated2}}
\end{subfigure} &
\begin{subfigure}{0.25\textwidth}
	\centering
	\includegraphics[trim = 0mm 0mm 0mm 0mm, clip, width=1\linewidth]{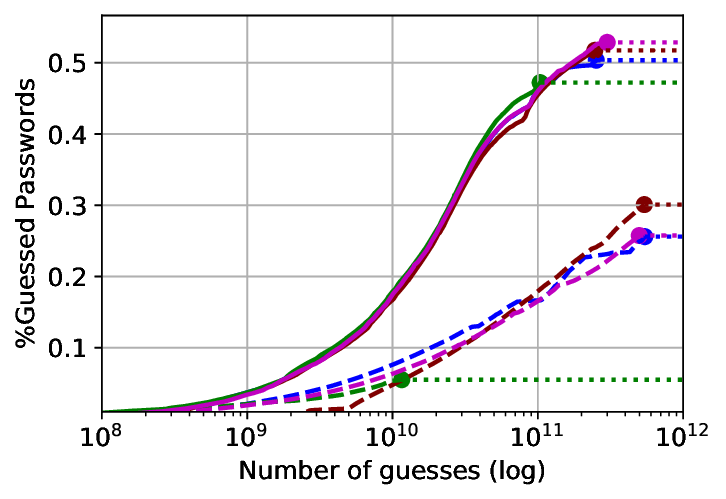}
	\vspace{-0.7cm}
	\caption{\textit{zooks} with \textit{generated2}}
\end{subfigure}&
\begin{subfigure}{0.25\textwidth}
	\centering
	\includegraphics[trim = 0mm 0mm 0mm 0mm, clip, width=1\linewidth]{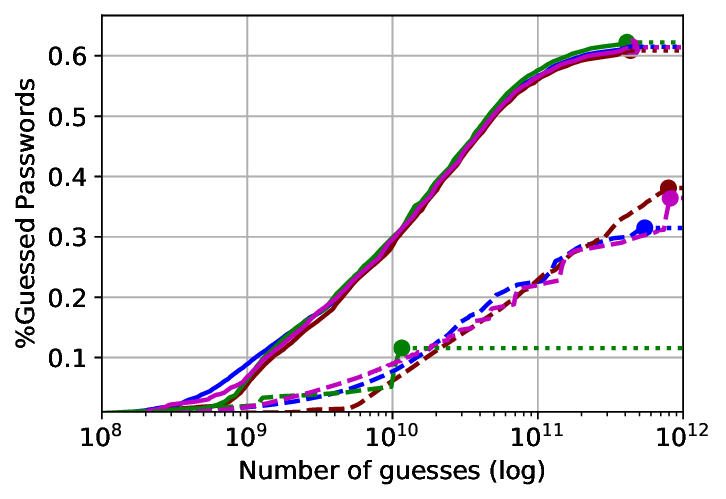}
	\vspace{-0.7cm}
	\caption{\textit{youku} with \textit{generated2}}
\end{subfigure}&
\begin{subfigure}{0.25\textwidth}
	\centering
	\includegraphics[trim = 0mm 0mm 0mm 0mm, clip, width=1\linewidth]{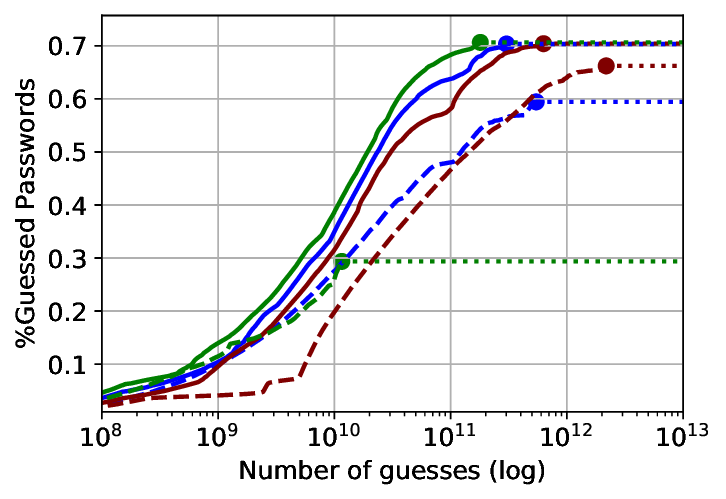}
	\vspace{-0.7cm}
	\caption{\textit{RockYou} with \textit{generated2}}
\end{subfigure}
			\end{tabular}
	}
	\caption{Each plot reports the number of guesses (in log scale) and the percentage of matched passwords for different rule-sets and dictionaries against several attacked-sets. Each row reports a rule-set, whereas each column identifies an attacked-set. We use four dictionaries, each identified by a colored line. Continuous lines show \adam~attacks whereas dashed lines refer to standard mangling rules attacks.}
	\label{fig:results}
\end{figure*}
	\section{Adaptive, Dynamic Mangling rules: \adam}
	\label{sec:adam}
The results of the previous section confirm the effectiveness of the dynamic guessing mechanisms. We increased the number of hits compared to classic dictionary attacks by using the produced guesses to improve the attack on the fly. However, in the process, we also increased the number of guesses, possibly in a way that is hard to control and gauge. Moreover, by changing the dictionary at runtime, we disrupt any form of optimization of the initial configuration, such as any \textit{a priori} ordering in the wordlist~\cite{reasoning} and any joint optimization with the rules-set\footnote{I.e., new words may not interact well with the mangling rules in use.}. Unavoidably, this leads to sub-optimal attacks that may overestimate passwords strength.
\par

To mitigate this phenomenon, we combine the dynamic augmentation technique with the complementary Adaptive Mangling Rules framework. The latter seeks an optimal configuration at runtime on the dynamic dictionary, promoting compatibility with the rules-set and limiting the impact of imperfect dictionary-words even if these are unknown before the attack. This process is further supported by the dynamic budgets that address the possible covariate-shift~\cite{covariate_shift} of the compatibility function induced by the augmented dictionary.
\par

Hereafter, we refer to this final guessing strategy as \adam~(\textbf{A}daptive, \textbf{D}yn\textbf{a}mic \textbf{M}angling rule\textbf{s}). Details on the implementation of \adam~are given in Appendix~\ref{app:imple}, whereas we benchmark it in Appendix~\ref{app:bench}.
\subsection{Evaluation}
 Figure~\ref{fig:results} reports an extensive comparison of \adam~against standard mangling-rules attacks. In the figure, we test all pairs of dictionary/rule-set obtained from the combination of the dictionaries: \textit{MyHeritage}, \textit{RockYou}, \textit{animoto}, \textit{phpBB} and the rules-sets: \textit{PasswordPro}, \textit{generated}  and  \textit{generated2} on four attacked-sets. 
  Hereafter, we switch to a logarithm scale given the heterogeneity of the number of guesses produced by the various configurations.
 \par 
For the reasons given in the previous sections, \adam~outperforms standard mangling rules within the same configurations, while requiring fewer guesses on average. More interestingly, \adam~attacks generally exceed the hits count of all the standard attacks regardless of the selected dictionary. In particular, this is always true for the \textit{generated} and \textit{generated2} rules-sets.
\par
 Conversely, in cases where the dynamic dictionary augmentation offers only a small gain in the number of hits (\eg attacking \textit{RockYou}), \adam~equalizes the performance of various dictionaries, typically, towards the best configuration for the standard attack. In Figures~\ref{fig:rrp} and~\ref{fig:rrg}, all the configurations of \adam~reach a number of hits comparable to the best configuration for the standard attack, i.e., using \textit{MyHeritage}, while requiring up to an order of magnitude fewer guesses (\eg Figure~\ref{fig:rrp}), further confirming that the best standard attack is far from being optimal. In the reported experiments, the only outlier is \textit{phpBB} when used against \textit{zooks} in Figure~\ref{fig:rzp}. Here, \adam~did not reach/exceed all the standard attacks in the number of hits despite consistently redressing the initial configuration. However, this discrepancy is canceled out when more mangling rules are considered such as in Figure~\ref{fig:rzg}.
 \par
 
{Eventually, the \adam~attack makes the initial selection of the dictionary systematically less influential.} For instance, in our experiments, a set such as \textit{phpBB} reaches the same performance of wordlists that are two orders of magnitude larger (\eg \textit{RockYou}). The crucial factor remains the rules-set's \textbf{cardinality} that ultimately determines the magnitude of the attack, even though it does not appreciably affect the guessing performance.
\par
 
\begin{figure}[t]

	\centering
		\includegraphics[trim = 35mm 93mm 25mm 0mm, clip, width=.4\linewidth]{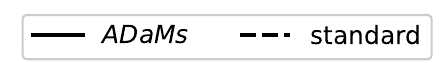}\\
\resizebox{.45\textwidth}{!}{
	\begin{subfigure}{.24\textwidth}
		\centering
		\includegraphics[trim = 0mm 0mm 0mm 0mm, clip, width=1\linewidth]{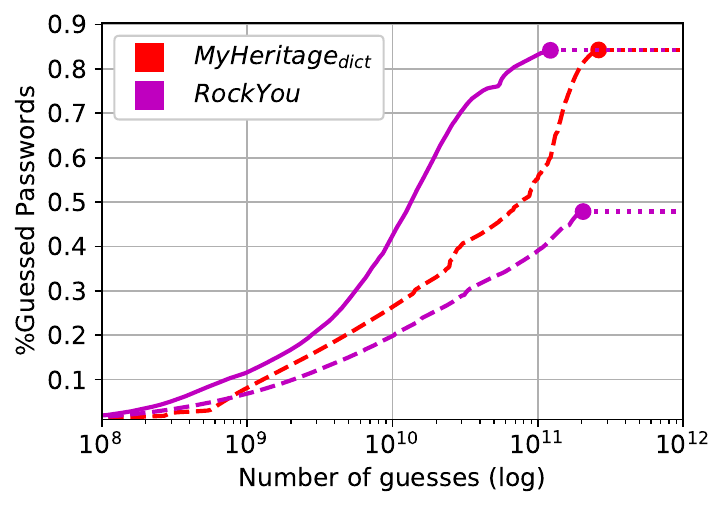}
		\caption{$\textit{MyHeritage}_{\text{target}}$}
	\end{subfigure}\begin{subfigure}{.24\textwidth}
		\centering
		\includegraphics[trim = 0mm 0mm 0mm 0mm, clip, width=1\linewidth]{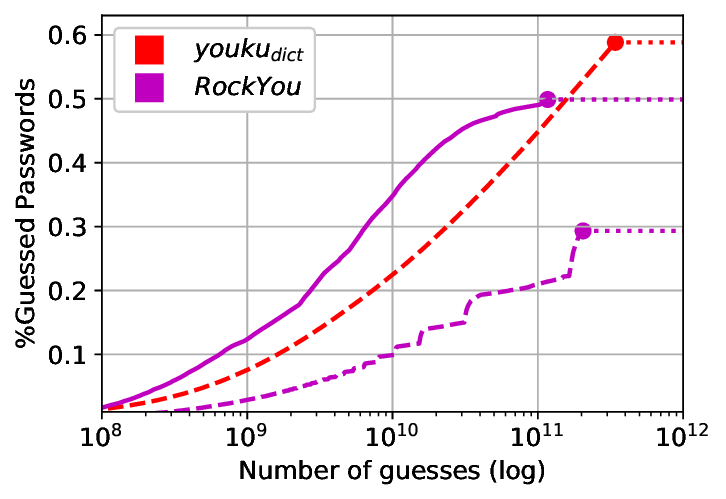}
		\caption{$\textit{youku}_{\text{target}}$}\label{fig}
	\end{subfigure}
}
	\caption{Comparison of \adam~against optimal dictionary for two sets of passwords.}
	\label{fig:optimal}
\end{figure}

The effectiveness of \adam~is better captured by the results reported in Figure~\ref{fig:optimal}. 
Here, we create a synthetic optimal dictionary for an attacked-set and evaluate the capability of \adam~to converge to the performance of such an optimal configuration. To this end, given a password leak $X$, we randomly divide it in two disjointed sets of equal size, say $X_{\text{dict}}$ and $X_{\text{target}}$. Then, we attack $X_{\text{target}}$ by using both $X_{\text{dict}}$ (\ie optimal dictionary) and an external dictionary (\ie sub-optimal dictionary). Arguably, $X_{\text{dict}}$ is the \textit{a priori} optimal dictionary to attack $X_{\text{target}}$ since $X_{\text{dict}}$ and $X_{\text{target}}$ are samples of the very same distribution.
\par

We report the results for two sets: \textit{MyHeritage} and \textit{youku}. The attacks are carried out by using the rules-set \textit{generated} and \textit{RockYou} as the external dictionary. 
In the case of \textit{MyHeritage}, the \adam~attack is more precise than the optimal dictionary and produces a comparable number of hits. Similarly, in the case of \textit{youku}, the \adam~attack guesses faster than the optimal dictionary within the first $10^{11}$ guesses. However, in this case, it does not reach an equivalent number of guessed passwords. We can attribute this to the high discrepancy between the initial dictionary \textit{RockYou} and the attacked-set \textit{youku} that cannot be bridged without prior knowledge.\footnote{The leak \textit{youku} is mostly composed of Chinese passwords that are underrepresented in \textit{RockYou}.}
Nevertheless, the dictionary augmentation technique can induce a dictionary that has a comparable utility to one of the best optimal \textit{a priori} setup, while requiring no information on the attacked-set. In the process, the adaptive framework consistently accounts for the noise introduced by the augmentation, allowing \adam~to be even more precise than the optimal dictionary for most of the attack (\ie within the first $10^{11}$ guesses).
\par

 Further comparison with other password models can be found in Appendix~\ref{app:comparison}. 

%
%
 %
%

	\section{Takeaways and New Directions}
	\label{conclusion}
	The \adam~attack autonomously pushes the attack strategy towards the optimal one, producing password strength estimates that better model actual adversarial capabilities. As shown in Figure~\ref{fig:results}, the approach also makes the guessing attack more resilient to deficiencies in the initial configuration, reducing the bias induced by misconfiguration. 
	In this direction, the \adam~attack further proves the intrinsic unsuitability of arbitrarily chosen configurations and the overestimation of password security that those can induce.
	\par
	
	Compared with other systems~\cite{imprl, FLA}, our framework provides researchers and security practitioners with a markedly more efficient and flexible solution. 
	We make our code and trained models publicly available\footnote{\url{https://github.com/TheAdamProject/adams}} in the hope {\bf our system will help improve the soundness of password strength estimation techniques}.
	\par
	Finally, our techniques pave the way for new valuable directions in the study of password security: \textbf{(1)}~our dynamic attack offers a framework capable of explaining causality relations among guessed passwords in a dynamic context; the \textit{hits-tree} produce from our technique could provide insights on how to proactively reduce the threat of dynamic attackers. \textbf{(2)}~Mangling rules are not necessarily \textit{effective} or \textit{ineffective} as assumed in current automatic configuration techniques~\cite{auto_mr,reasoning}. They have a conditional nature that must be accounted for to seek optimal configurations. Adaptive mangling rules have proven to be superior and more effective. Still, it would be interesting to devise new techniques to automatically formulate mangling rules rather than select and compose existing ones.
	\section*{Acknowledgements}
	We wish to thank Blase Ur (our {\em shepherd}) and the other anonymous reviewers for the valuable feedback which helped to improve the paper.
	\bibliographystyle{plain}
	\bibliography{bib}
	\section*{Appendices}
\setcounter{section}{0}
\counterwithin{table}{section}
\counterwithin{figure}{section}
\renewcommand{\thesection}{\Alph{section}}%

\section{Comparison with other password models}
\label{app:comparison}
\begin{figure*}[t]
	\centering
	\includegraphics[trim = 0mm 87mm 0mm 0mm, clip, width=.6\linewidth]{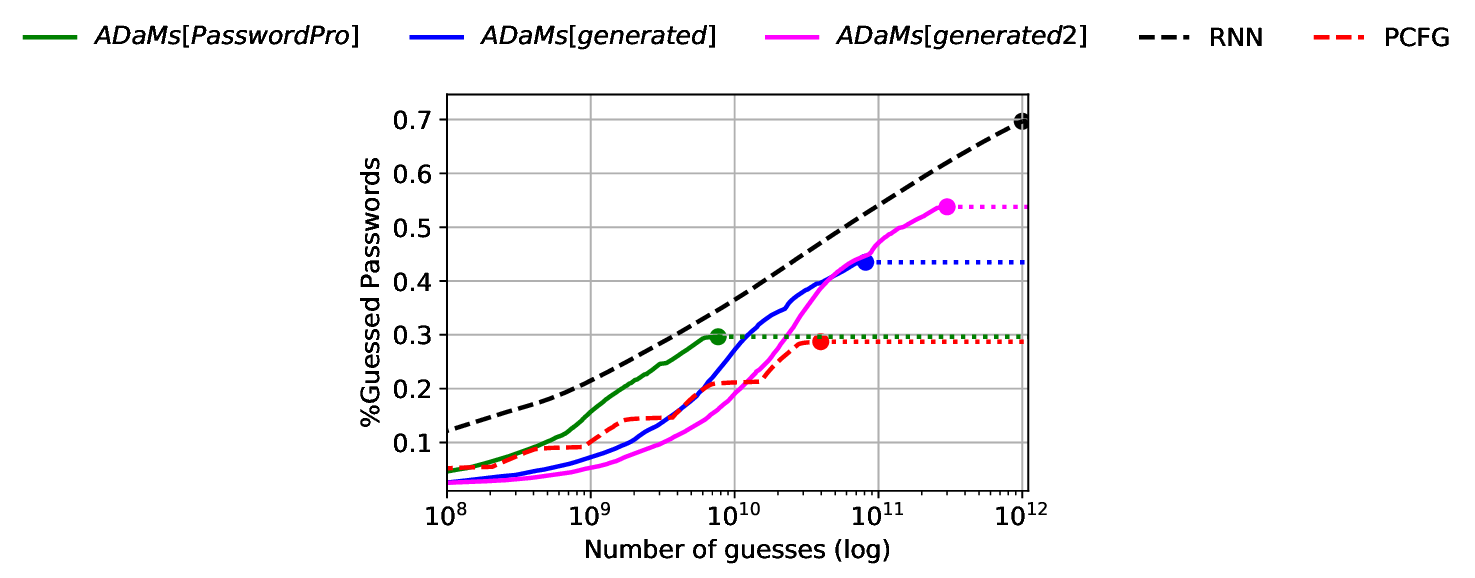}\\
	\begin{subfigure}{.27\textwidth}
		\centering
		\includegraphics[trim = 0mm 0mm 0mm 0mm, clip, width=1\linewidth]{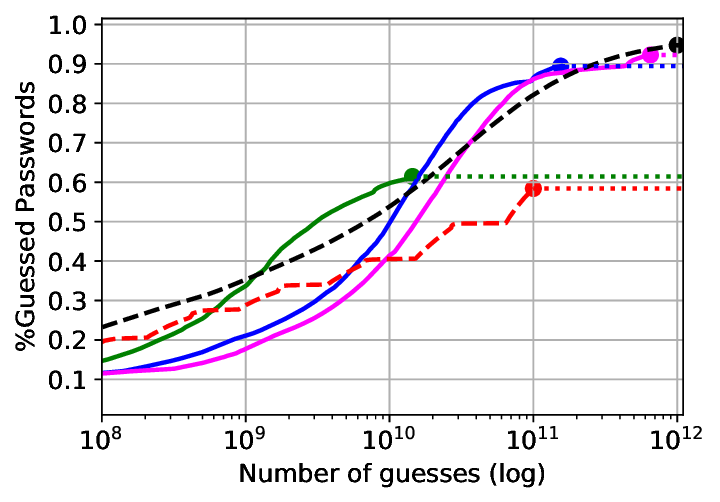}
		\caption{\textit{MyHeritage}}\label{}
	\end{subfigure}\begin{subfigure}{.27\textwidth}
		\centering
		\includegraphics[trim = 0mm 0mm 0mm 0mm, clip, width=1\linewidth]{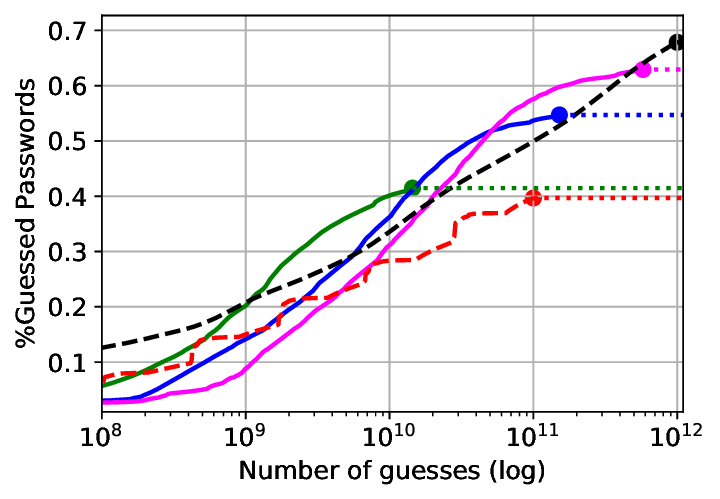}
		\caption{\textit{youku}}\label{fig}
	\end{subfigure}\begin{subfigure}{.27\textwidth}
		\centering
		\includegraphics[trim = 0mm 0mm 0mm 0mm, clip, width=1\linewidth]{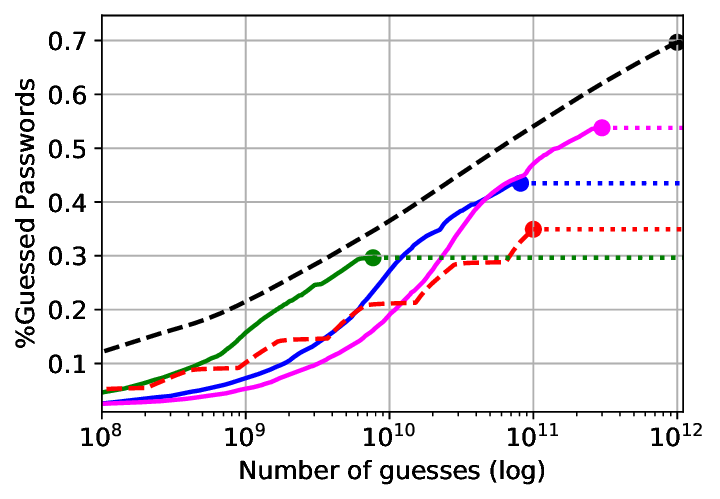}
		\caption{\textit{zooks}}\label{fig}
	\end{subfigure}
	\caption{Comparison of the \adam~attacks against the RNN-based approach of Melicher~\etal~\cite{FLA} and PCFG~\cite{PCFG} for three password leaks.}
	\label{fig:vs_flal}
\end{figure*}
Next, we compare \adam~with other password models.
\par

Figure \ref{fig:vs_flal} reports a direct comparison against the RNN-based approach of Melicher~\etal~\cite{FLA} and PCFG~\cite{PCFG}. The RNN-based password model is the state-of-the-art for password strength estimation, although its computational cost in generating guesses makes it impractical for real password guessing. We train the model using \textit{RockYou} and simulate password guessing attacks using~\cite{monte}. In the process, we use default parameters of the available software~\cite{FLAgit} and consider passwords with guess-number within $10^{12}$.\\
PCFG is the academic approach that better mirrors the guessing generation process of dictionary attacks. We train the PCFG-based model on \textit{RockYou} using the default setting~\cite{PCFGgit}. In this case, we limit to the first $10^{11}$.
\par

We compare the models on three leaks: \textit{MyHeritage}, \textit{youku} and \textit{zooks}. For the \adam~attacks, we use \textit{RockYou} as a dictionary, whereas we report results for three rules-sets.\\
Surprisingly, the \adam reach performance very close to the one obtained from the RNN-based model. It even outperforms the parametric attack in two of the three attack-sets. Similarly, \adam~tend to perform better than PCFG in the three cases, especially after the initial guesses.
\begin{figure}[h]
	\centering
	\includegraphics[trim = 0mm 78mm 0mm 0mm, clip, width=.8\linewidth]{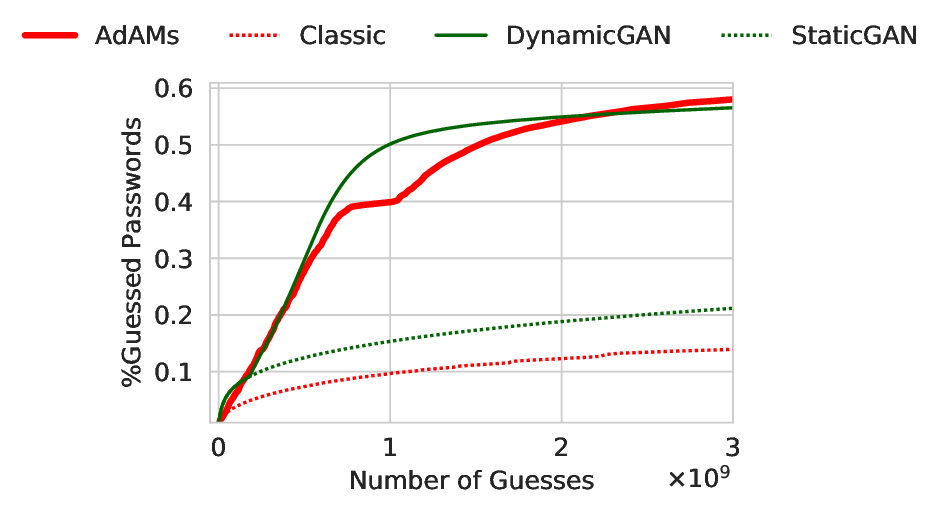}\\
	\begin{subfigure}{.23\textwidth}
		\centering
		\includegraphics[trim = 0mm 0mm 0mm 0mm, clip, width=1\linewidth]{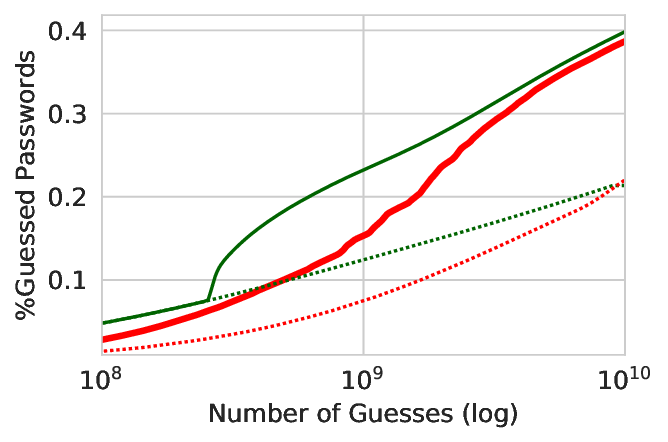}
		\caption{\textit{youku}}
	\end{subfigure}\begin{subfigure}{.23\textwidth}
		\centering
		\includegraphics[trim = 0mm 0mm 0mm 0mm, clip, width=1\linewidth]{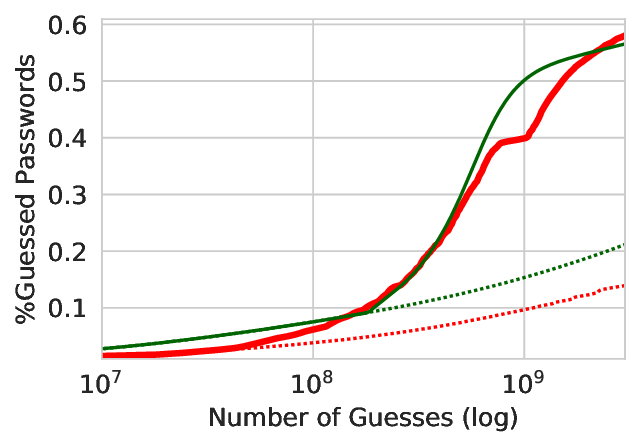}
		\caption{\textit{zomato}}\label{fig}
	\end{subfigure}
	\caption{Performance comparison between \adam~and the dynamic attack~\cite{imprl}. Classic mangling rules attacks and StaticGAN~\cite{imprl} are reported as baseline.}
	\label{fig:vs_gan}
\end{figure}
Furthermore, Figure~\ref{fig:vs_gan} compares \adam~against the original GAN-based dynamic attack~\cite{imprl}. We base the comparison on the same leaks used in~\cite{imprl}; namely, the \textit{youku} and \textit{zomato} leak (details given in Table~\ref{tab:password}). The GAN-based model is trained on the \textit{RockYou} leak and the attack is performed with the same hyper-parameters used in~\cite{imprl}: $\sigma=0.35$ and \textit{hot-start} $\alpha=10\%$. Despite our simpler approach, the \adam~attack performs very similarly to the GAN-based attack, besides being significantly faster in generating guesses (see Table~\ref{tab:bench}).
%
%
%
\section{Details on the deep learning framework}
\label{app:arch}
\begin{algorithm}[h]
	\footnotesize
	\KwData{input tensor: $x_{in}$}
	$x = \texttt{batchNormalization}(x_{in})$\;
	$x = \texttt{ReLU}(x)$\;
	$x = \texttt{1D-Convolution}(x, f, k)$\;
	$x = \texttt{batchNormalization}(x)$\;
	$x = \texttt{ReLU}(x)$\;
	$x = \texttt{1D-Convolution}(x, f, k)$\;
	\Return $ x_{in} + 0.3\cdot x $
	\caption{Residual Block: \texttt{residualBlock($\cdot$)}:}
	\label{algo:rb}
\end{algorithm}\begin{algorithm}[h]
\footnotesize
	\KwData{input tensor: $x_{in}$, rules-set $R$}
	$x = \texttt{charactersEmbedding}(x_{in}, 128)$\;
	$x = \texttt{1D-Convolution}(x, f, k)$\;
	\For{$0\ \textbf{to}\ d$}{
		$x = \texttt{residualBlock(x)}$
	}
	$bneck = \lceil \frac{f}{b} \rceil$\;
	$x = \texttt{1D-Convolution}(x, bneck, k)$\;
	$x = \texttt{flattern}(x)$\;
	$logits = \texttt{dense}(x, |R|)$\;
	\Return $logits$
	\caption{Architecture:}
	\label{algo:arch}
\end{algorithm}
This Appendix details the architecture used to implement the neural approximations of the compatibility functions presented in Section~\ref{sec:learning_aff}. It can be defined using five parameters, namely:
\begin{itemize}
	\itemsep0em 
	\item \textbf{Depth ($d$):} The number of residual blocks composing the network. Each residual block includes two 1D-convolutional layers, supported by normalization layers and activation \ie Algorithm~\ref{algo:rb}.
	\item \textbf{Number of filters ($f$):} The number of filters for each convolutional layer in the network.
	\item \textbf{Kernel size ($k$):} Size of the kernel used in every convolutional layer in the network.
	\item \textbf{Final Bottleneck ($b$):} Reduction of the number of filters before the final dense layer.
\end{itemize} 
The final architecture is described in Algorithm~\ref{algo:arch}. Our biggest models are realizations of the parameters: $d\myeq15,\ f\myeq512,\ k\myeq5$. We use $b\myeq2$ for \textit{PasswordPro} and \textit{generated}, $b\myeq3$ for \textit{generated2} instead.
%
%
%
%
\section{Impact of the Dynamic budget}
\label{app:dynbud}
\begin{figure}[t]
	\centering
	\includegraphics[trim = 0mm 95mm 0mm 0mm, clip, width=.7\linewidth]{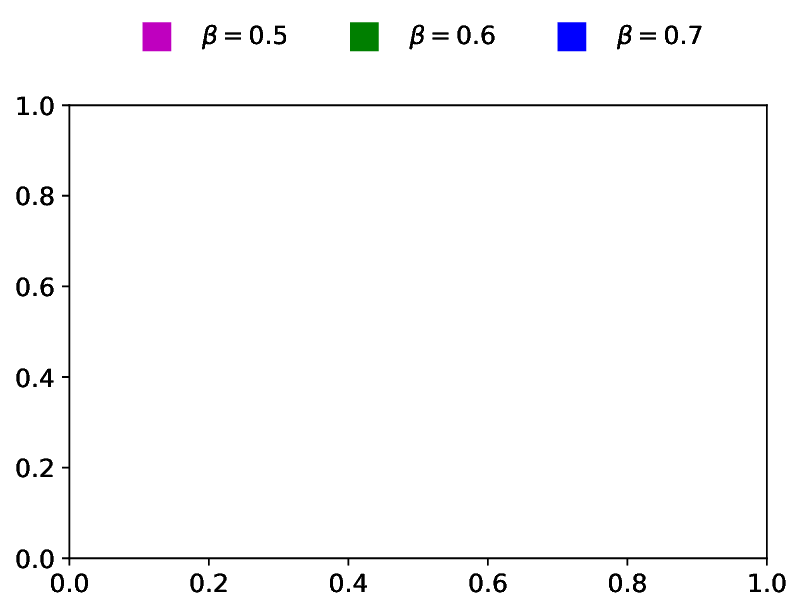}\\
	\centering
	\includegraphics[trim = 0mm 0mm 0mm 0mm, clip, width=.8\linewidth]{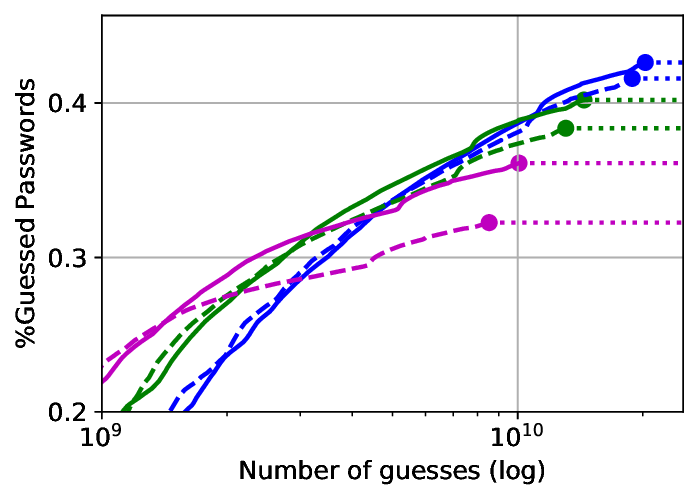}
	\caption{Effectiveness of the dynamic-budget within~\adam~for different value of $\beta$. Continuous lines present \adam, whereas dashed lines are \adam~ablated of the dynamic-budget}
	\label{fig:dynbud}
\end{figure}
We briefly illustrate the impact of the dynamic budget (\ie Section~\ref{sec:dyn_bud}) on the performance of \adam. As previously discussed, the dynamic budget has always a positive or neutral effect. Figure~\ref{fig:dynbud} reports an example for the attacked-set \textit{youku}. In the figure, continuous lines refer to the complete \adam~attack, whereas dashed lines report the results for \adam~without dynamic budget for the same configuration. We report the results for three values of $\beta$.\\
As shown in the example, the dynamic budget is particularly effective when low $\beta$ is used. In these cases, the dynamic logic helps better organize the small total budget of the attack, resulting in better global performance. The gain decreases when bigger budgets are adopted.
\section{Benchmarks}
\label{app:bench}
In this Appendix, we analyze the computational cost of generating guesses with \adam. 
Primarily, we test the overhead with respect to standard mangling rules (\ie \textit{Hashcat} \textit{CPU legacy}). 
\par
\begin{table}[t]
	\centering
	\caption{Number of guesses per second compute single core/GPU on a NVIDIA DGX-2 machine.}
	\label{tab:bench}
	\resizebox{.9\columnwidth}{!}{%
		\label{table:bench}
		\begin{tabular}{c|c|c|c|c}
			\toprule
			\makecell{ \adam\\\textbf{generated2}} & \makecell{ \adam\\\textbf{generated}}& \makecell{ \adam\\\textbf{PasswordPro}} & \makecell{ \textbf{Hashcat} \\ CPU legacy } & \makecell{ \textbf{GAN~\cite{imprl}}\\ Dynamic Attack}\\ \midrule
			$726182$ g/s & $709439$ g/s & $644444$ g/s & $928647$ g/s & $34189$ g/s\\
			\bottomrule
		\end{tabular}
	}
\end{table}
For the comparison, we produce $10^9$ strings and compute the number of guesses generated per second (\ie $g/s$). \textbf{In the process, we include the time of checking for the guesses in the set of the attacked passwords} (the same methodology is used for each tool and may not be computationally optimal). Note that we do not perform any hash function computation in the process.
We repeat the test $5$ times using \textit{RockYou} as dictionary and \textit{animoto} as attacked-set, whereas we repeat for the rules-sets: \textit{PasswordPro}, \textit{generated} and \textit{generated2}. Table~\ref{table:bench} averages the time for each tool. The result for the standard mangling rules is reported as average over the three rules-sets. Additionally, we report the timings for the GAN-based, dynamic attack described in~\cite{imprl}.
\par

On average, \adam~are just $25\%$ slower than standard mangling rules. Considering that the Adaptive mangling rules can reduce the number of guesses up to an order of magnitude, this overhead becomes negligible in practice. Moreover, this discrepancy easily fades out when slow hash functions, such as \cite{pkcs,bcrypt,mem_hard}, are considered.
\section{Implementation of \adam}
\label{app:imple}
We rely on the \textit{CPU legacy} version of \textit{Hashcat}\footnote{\url{https://github.com/hashcat/hashcat-legacy}} to implement \adam~attacks. Our prototype uses the CPU version as it is easier to modify its workflow, although the \textit{Hashcat} GPU engine can trivially support our approach.\footnote{The GPU engine is also more suited as it would naturally support the computation of the neural network on GPU, removing the CPU/GPU communication overhead}\\
In the code, we modify the main loop of \textit{Hashcat}, where it scans over dictionary words and then iterates on all rules. We read a batch of words from the dictionary, we give them as input to the neural network, and then, for each word $w$ in the batch, we apply only the rules whose values of $\alpha_R$ are greater than $(1-\beta)$. We check all these guesses and, those who match are added on top of the remaining words in the dictionary, \ie they will be part of the next batch of words. The same batching approach is used for the dynamic budget. Here, budget increments and normalization per rule are performed conjointly after every batch to further reduce computational overhead. In the implementation, we use batch-size equals to $4096$ dictionary words.
\section{Visualizing the compatibility function}
\begin{figure*}[t]
	\centering
	\resizebox{1\textwidth}{!}{%
		\includegraphics[scale=1]{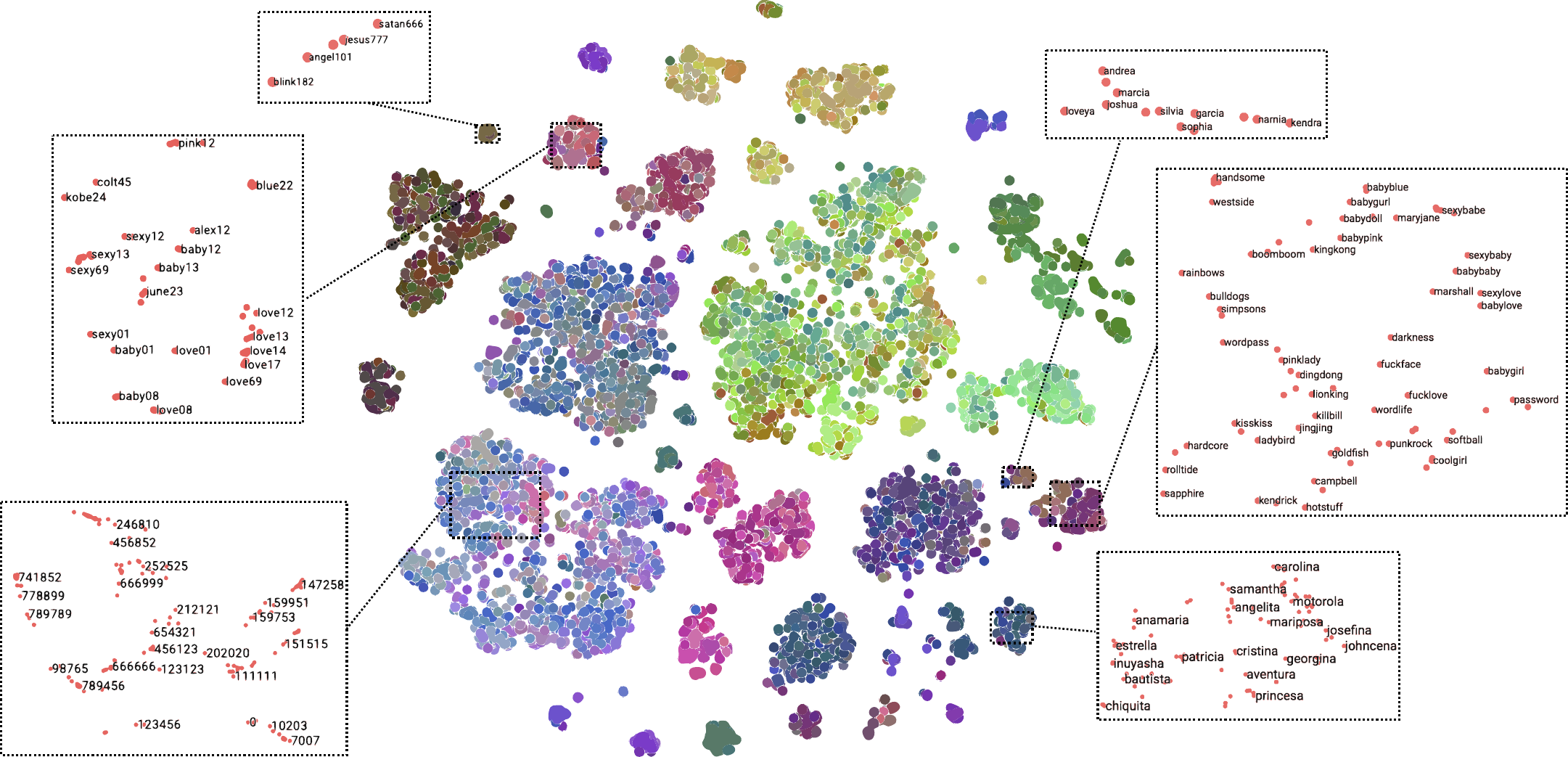}
	}
	\caption{Two-dimensional visualization of the dictionary-space representation learned from a model trained on the rule-set \textit{best64} and \textit{LinkedIn}. The visualization is obtained by projection into a two-dimensional space (the output of the last residual block via the \textit{T-SNE} algorithm). Reported colors represent the rules activated from the corresponding dictionary-word. Those are obtained by mapping the label-space to a three-dimensional space via \textit{T-SNE} algorithm. Successively, the latter is mapped into the \textit{RGB} domain, achieving visible colors.}
	\label{fig:rep}
\end{figure*}
\label{app:viz}
The dictionary-space representation learned from the neural net during the training provides a valid intuition over the \textit{compatibility} concept formulated earlier in the paper. Indeed, just visualizing it, we can make explicit many of the core assumptions we used to build Section~\ref{sec:fcd}.
\par
Figure \ref{fig:rep} reports a two-dimensional depiction of such dictionary-space representation obtained using a small set of strings sampled from the \textit{RockYou} password leak. In the figure, every point represents a dictionary-word. Each point's color indicates which rules are activated from the corresponding string in consideration of $\Xgt$. These colors are arranged so that similar colors map to sets of active rules with a large intersection.\\
As shown in the figure, the representation learned from the model organizes dictionary-words that activate the same/similar set of rules close to each other in the space, making explicit sets of clusters that partition the dictionary-space. Intuitively, each of these clusters describes a semantic partition of the dictionary-space, dedicated to activating just a specific subset of mangling rules. Going back to the intuition of Section \ref{sec:fcd}, such clusters can be seen as the activation domains of the corresponding mangling rules.\\
Furthermore, the semantic segmentation phenomenon can also be explained by observing samples from the various clusters. These are reported enclosed by dashed rectangles in Figure \ref{sec:fcd}. In general, in the representation learned by the model, larger clusters group macro classes of dictionary-words, such as purely numeric or purely alphabetic strings.\footnote{As easily predictable, the mangling rules that activate these macro classes of dictionary-words tend to not overlap.}. Smaller clusters, instead, tend to bound more specific patterns. For example, clusters on the picture's bottom right indicate how the representation binds words representing first names. It is important to note that no semantic features are given to the model. The model learns this semantic relation solely from dictionary-words' functional meaning; that is, two strings are \TT{similar} as they activate the same rules.
\end{document}